
\baselineskip=0.6cm   
\font\main=cmr10 scaled \magstep1
\font\sub=cmr10 scaled \magstep0
\hsize  = 6.5 truein
\vsize  = 9.3 truein
\nopagenumbers
\headline={\hss -- \folio\ -- \hss}
\raggedbottom
\def\hangpara{\par\hangindent 25pt\noindent}
\def\Lya{Ly$\alpha$ }
\def\Lyb{Ly$\beta$ }
\def\Lyg{Ly$\gamma$ }
\def\Lyd{Ly$\delta$ }
\def\etal{et al. }
\def\kms{km~s~$^{-1}$ }
\def\L{$\lambda$}
\def\cm-2{cm$^{-2}$}
%
%
%
\main
\centerline{}
\vskip 1.0truein
\centerline{\bf LICK OPTICAL SPECTRA OF}
\centerline{\bf QUASAR HS~1946+7658 AT 10 km/sec}
\centerline{\bf RESOLUTION: LYMAN-ALPHA FOREST AND}
\centerline{{\bf METAL ABSORPTION SYSTEMS}~\footnote{\raise3pt\hbox{1}}
{Based on observations obtained at Lick Observatory, University of
California.}}
\bigskip
\bigskip
\centerline{XIAO-MING FAN\footnote{\raise3pt\hbox{2}}
{Center for Astrophysics and Space Sciences, 0111, University of California,
San Diego, La Jolla, CA 92093--0111. (fan@cass154.ucsd.edu,
tytler@cass155.ucsd.edu)}{\raise3pt\hbox{,}}
\footnote{\raise3pt\hbox{3}}{Department of Astronomy, Columbia University,
New York}
and DAVID TYTLER{\raise3pt\hbox{2}}{\raise3pt\hbox{,}}
\footnote{\raise3pt\hbox{4}}
{Department of Physics, University of California, San Diego}
}
\vskip 0.50truein
\centerline{$\underline{To~appear~in~the~Astrophysical~Journal~Supplements}$}
\bigskip\bigskip
\centerline{Received: \hbox to 1.5truein{$\underline{January~14,~1994}$}}
\centerline{Accepted: \hbox to 1.5truein{$\underline{February~10,~1994}$}}
\vfill\eject
%
%
\centerline{}
\centerline{ABSTRACT}
\medskip

We present optical spectra of the most luminous known QSO HS~1946+7658
($z_{em}=3.051$).  Our spectra have both full wavelength coverage,
3240 -- 10570~\AA\ , and in selected regions, either high signal to
noise ratio, SNR $\simeq$ 40 -- 100, or unusually high
$\sim 10 \rm ~km~s^{-1}$ resolution, and in parts of the \Lya forest and
to the red of \Lya emission they are among the best published.  We find
113 \Lya systems and six metal line systems, three of which are new. The
metal systems at $z_{abs}=2.844$ and 3.050  have complex velocity structure
with four and three prominent components respectively.

We find that the system at $z_{abs}=2.844$ is a damped \Lya absorption
(DLA) system,
with a neutral hydrogen column density of log $N$(H~I) = 20.2$\pm$0.4,
and it is the cause of the Lyman limit break at \L $\sim$ 3520~\AA.
We believe that most of the H~I column density in this system is in
$z_{abs}=2.8443$ component which shows the strongest
low ionization absorption lines.  The metal abundance in the gas
phase of the system  is [M/H]$\simeq -2.6\pm0.3$, with a best estimate
of [M/H]=$-2.8$, with ionization parameter $\Gamma=-2.75$, from a
photoionization model. The ratios of the abundances of C, O, Al, and Si
are all within a factor of two of solar, which is important for two
reasons.  First, we believe that the gas abundances which we measure are
close to the total abundances, because the ratio of Aluminum to other
elements is near cosmic, and Al is a refractory element which depletes
very readily, like Chromium, in the interstellar medium.  Second, we do
not see the enhancement of O with respect to C of [O/C] $\simeq 0.5 - 0.9$
reported in three partial Lyman limit systems by Reimers \etal 1992 and
Vogel \& Reimers 1993; we measure [O/C] $=-0.06$ for observed ions and
[O/C] $\simeq 0.2$ after ionization corrections, which is consistent with
solar abundances.  We see C~II$^*$(\L1335) offset by 15 \kms with
respect to  C~II(\L1334), presumably because the gas density varies from
2 -- 8 cm$^{-3}$ with changing velocity in the DLA system.  These densities
imply that the damped component is 6 -- 25 pc thick, which is reasonable
for a single cloud in a cold spiral disk.

The system at $z_{abs}=1.7382$ is also believed to be damped with
$N$(H~I) $\sim 10^{21}$ \cm-2, because we see Cr~II, but its \Lya line
will never be seen because it is below the Lyman limit system (LLS)
in the other DLA system.

We see a 2.6$\sigma $ lack of \Lya forest lines well away from the QSO
redshift, which may be a chance fluctuation. We also see a correlation
between column density $N$(H~I) and Doppler parameter $b$
for 96 unsaturated \Lya forest absorption lines, and although this
correlation persists in the 36 \Lya lines which lie in regions where
the SNR $\simeq$ 8 -- 16, we agree with Rauch \etal (1993) that it
is probably a bogus effect of low SNR.
The same applies to lines with very low $b$ values: in regions where
SNR $\leq 8$ we see many \Lya lines which appear to have
$10 \leq b \leq 20$, but
when $8 \leq {\rm SNR} \leq 16$ we see only one line with $b \leq 15$ \kms, and
two others which we believe have $b \leq 20$, with values of 20 and 16~\kms.
Traditional \Lya line samples which include all lines which have
$W/\sigma (W) \geq 4$ are not adequate to explore the distribution of
the properties of individual clouds, because we need much
higher $W/\sigma (W)$ and SNR to avoid the strong biases.
\bigskip
\noindent
{\it Subject Headings:} cosmology -- galaxy: intergalactic medium
-- galaxy: abundance -- quasars: absorption lines
-- quasars: individual : HS~1946+7658

\vfill\eject
%
%
\centerline{1. INTRODUCTION}
\medskip

QSOs at high redshift are interesting because they existed long time ago,
and their optical spectra contain several absorption systems with metal
lines and many hundreds of Lyman-alpha (\Lya) forest absorption lines.
The metal
systems arise from high redshift galaxies, which are usually too faint to
see. We can measure many of the properties of
their absorbing gas, including the ionization, chemical
composition and velocity distribution, but usually not the density or size,
and we want to know what these absorbers can tell us about young galaxies.

The main issues with the high redshift \Lya forest lines, are their column
densities, ionization, metallicity, velocity dispersions, clustering
and origins. We originally proposed that high redshift \Lya systems
are intergalactic clouds (Sargent \etal 1980), but
our recent work on clouds at low redshifts $z \leq 1$ have shown
that at least $35 \pm 10$\%, and probably $65 \pm 18$\% of these \Lya forest
systems arise from the outer regions of normal galaxies (Lanzetta \etal 1994).

Studies of both the metal lines and the \Lya forest lines require
high resolution spectra, because we need to resolve the absorption
lines, which are typically 10 -- 40 \kms wide, to get reliable column
densities. Unfortunately only a hand full of the 350 known QSOs at
$z_{em} \geq 2.8$ are bright enough for echelle spectra with 4-m
telescopes.

Here we present FWHM=10 \kms echelle spectra of QSO HS~1946+7658,
which is the most luminous known QSO, with V=15.85 (Hagen \etal 1992)
and $z_{em}$=3.051 (our new value). Prior to Keck,
only two other QSOs have been
observed at higher resolution (Q2206$-$199N at FWHM=6 \kms by Pettini
\etal 1990a; and Q1100$-$264 at FWHM=8.5--9 \kms by Carswell \etal 1991),
and a few others at lower resolutions (Q2126$-$158 at
FWHM=14 \kms by Giallongo \etal 1993; Q0014+813 at FWHM=20 \kms by
Rauch \etal 1992; Q0420$-$388 at
FWHM$\sim$33 \kms by Atwood \etal 1985; Q2000$-$330 at FWHM$\sim$33 \kms by
Carswell \etal 1987; HS~1700+6416 at FWHM$\sim$35 \kms by  Sanz \etal 1993).
Our spectra have the highest signal to noise ratio per \AA ngstrom redward
of \Lya emission line, where metal lines predominate.

In this paper we present new data on the \Lya forest and the
metal systems.

\medskip
\centerline{1.1 {\sl  Lyman-alpha Forest}}
\medskip

Our first goal is the study of the column densities, velocity dispersions
and redshift distribution of the \Lya forest absorption lines.
Pettini \etal (1990a, hereafter PHSM) claimed
that most \Lya forest clouds have Doppler parameter
$b \leq 22$ \kms, and that $b$ correlates with column density $N$(H~I).
They interpreted their results to imply that (1) $b$ values are largely
determined by bulk rather than thermal motions, and hence (2) the \Lya
forest clouds may be mostly cold (T $\leq$ 5000 -- 10,000~K), thin,
dense filaments or sheets. Their results imply that the typical
\Lya forest cloud is 10$^{-5}$pc thick, which is uncomfortably low
and contradicts
the usual belief that the clouds are highly ionized with
temperatures of T $\sim 3\times10^4$~K and typical sizes of 10--100 kpc
(Sargent \etal 1980). This issue has been discussed in several recent
papers.
Carswell \etal (1991) and Rauch \etal (1992) did not see the $N$(H~I)--$b$
correlation in Q1100$-$264 and Q0014+813 respectively. Rauch \etal (1993)
re-analyzed the original PHSM spectrum of Q2206$-$199N and confirmed
the appearance of both the $N$(H~I) -- $b$ correlation, and the very
narrow lines, but their simulations showed that both
were probably artifacts of low SNR.

\medskip
\centerline{1.2 {\sl  Metal Abundances}}
\medskip

Our second goal is the study of the velocity structure and especially
the metal abundances of the metal line systems. There are six main
obstacles on the path to metal abundances.

\smallskip {\bf 1. Resolution of  the velocity components.}
The metal line systems nearly always consist of many different velocity
components, which produce strongly blended lines. For example, Lanzetta
\& Bowen (1992) presented FWHM=7--35 \kms spectra of four metal systems
at $0.4 < z_{abs} < 0.7$ from known galaxies. Two of these
are absorption complexes, with multi-component structure which covers
100 -- 300 \kms. This structure,
and especially the very narrow components with FWHM
$\leq$ 20 \kms, makes it hard to be certain that all of the gas has been
seen, which is the second obstacle.

\smallskip {\bf 2. Hidden low $b$ components.}
The components with the highest column density are often the coolest,
with the lowest ionization, so they have the smallest $b$ values, and
their absorption lines can be weak, and be hidden by other components.
The solution is to look at lines with lower and lower oscillator
strengths, until only one component remains visible, which will be the
one with the highest column density. This requires very high resolution
and SNR.

\smallskip {\bf 3. Column Densities for each component of each ion.}
Each velocity component in a system can have its own ionization,
abundance, and depletion, so we should try to get each of these properties
for each velocity component.  The column densities are clearly the key,
because if we know them for each component we can treat each as a
separate system, but they are hard to get.
For example, with hydrogen we have only
the Lyman series lines, and they all have relatively large oscillator
strengths so the components blend together. There are three solutions:
(1) use the damping profile of \Lya line to get the total column density,
assume that most of this H~I is associated with the component with
the most neutral gas and ignore all other components;
(2) use 21~cm absorption to resolve the H~I components, but only the
largest column densities absorb at 21~cm and there are few high
redshift radio loud QSOs;
(3) find systems with simple velocity structure, which might introduce
new biases.

\smallskip {\bf 4. Ionization.}
If all of the gas is neutral, then the observed column densities of the
low ions are the total column densities. But if it is ionized, then we
need to observe all abundant stages of ionization and sum their column
densities to get the total gas abundance.
We frequently need an ionization model to estimate the column densities
of ions which are not seen, as we will do here in \S 4.2.5 for a damped
\Lya absorption (hereafter DLA) system.

\smallskip {\bf 5. Depletion.}
There is now good evidence that significant proportions of the
refractory elements in some DLA
systems are in dust. Fall \& Pei (1989) have clearly shown that QSOs
with DLA systems have redder continua, which implies typical dust to
gas ratios of 5 -- 25\% of that in the Milky Way.

Absorption lines give {\it gas phase abundances}, which can be very
different from total abundances, because in the local ISM up to 99\%
of refractory elements like Cr, Al, Ca, Ni and Ti are in dust, and
only 1\% remain in the gas.
The Fall \& Pei results then imply that 4 -- 20\% of the
refractory elements will be in the gas phase.

The level of depletion can be estimated from the ratios of the
abundances of refractory to non-refractory elements.  Rauch \etal
(1990) found that the metal abundance in a DLA system at
$z_{abs}=2.076$ toward QSO 2206$-$199N is very low
[Si/H] $\le -2.4$, without much depletion onto dust.
\footnote{\raise3pt\hbox{5}}{We use the usual
notation [M/H] $\equiv$ log (M/H) $-$ log (M/H)$_{\odot}$
to compare measured abundance with solar values.}

Pettini \etal (1990b)  advocated the Cr/Zn column
density ratio because (1) in our Galaxy Cr depletes readily
while Zn does not, (2) both elements have lines with oscillator
strengths which are low, and we hope low enough that only
the component with the largest column density will be seen,
and (3) the lines have similar rest wavelength's.
Recently Pettini \etal (1994) measured Cr/Zn ratios in seventeen
DLA systems. They found gas phase abundances of about 0.1 solar for Zn, and
depletions which left 16\% of the Cr in the gas phase, compared with
1\% the local ISM,
which is consistent with the implication of Fall \& Pei's results.

\smallskip {\bf 6. Interpretation.}
There are two main problems of interpretation. First we do not know where
the absorption is occurring, and we expect abundances to vary both from
galaxy to galaxy and within a galaxy at a given epoch. Second, the
abundances can be weighted many ways: for example, we can
(1) take only the highest abundance in a system,
(2) average all components in a system, which may be one galaxy, or a
group of galaxies,
(3) calculate the cosmological average abundance, which is the mean
cosmological density of metals divided by that of hydrogen.

\bigskip

The partial Lyman limit systems and the DLA systems are best suited for
abundance work. The LLSs include all
absorption systems with sufficient H~I to be optically thick to
the Lyman continuum and they include all DLA systems. Some LLSs have
neutral hydrogen column densities 4 -- 5 orders of magnitude below
those of the DLA systems, but the total hydrogen column
densities can be comparable. It is very hard to get abundances for
ordinary LLS with $\tau \gg 1$ because no QSO
radiation is visible at \L $<$ 912~\AA\ and
there are not enough ions with lines at \L $>$ 912~\AA\
to show the level of ionization.
However there are numerous partial LLSs with $\tau \leq 1$
which do transmit radiation at \L $<$ 912~\AA, and both the
ionization and abundances can be measured in these systems using
the numerous metal absorption lines at $500 \leq \lambda \leq 900$~\AA.

The DLA systems are simpler. For them we normally assume that the
velocity component with the strongest low ionization lines is
completely neutral, and we ignore gas in all other velocity
components, much of which is ionized.

We already know that the LLSs have a different abundance
distribution from the DLA systems, because some LLSs are apparently
metal free, but all DLA systems show metal lines.

Early work on LLSs abundances includes Steidel (1990), who found
typical gas phase metal abundances in eight LLSs at
$z_{abs} \approx 3$ of [M/H] $\sim -2.5$, and Sargent \etal (1990)
who reported [M/H] $\leq$ $-$2.3 in one partial LLS at
$z_{abs}$=2.9676 toward QSO PKS 2126$-$158.
In an important set of papers, Reimers \etal (1992) and
Vogel \& Reimers (1993) showed that far UV lines allow easy
estimation of the ionization and hence abundances in partial LLS.
For three systems at $1.8 \leq z_{abs} \leq 2.4$ they find abundances
[C/H] $=-1.7, -1.9, -2.3$, and they see that both O and N are over
abundant with respect to C: [O/C] $=0.6, 0.8, 0.9$,
and [N/C] $=0.4, 0.8, 0.8$.

\medskip
\centerline{1.3 {\sl  Contents of This Paper}}
\medskip

We describe the spectrograph and the data reduction in \S2 and \S3.
In \S4 we discuss individual metal line absorption system, and we
estimate the gas phase metal abundance for the DLA system at
$z_{abs}$=2.8443. In \S5 we discuss the column density $N$(H~I), Doppler
parameter $b$ distribution and the $N$(H~I) - $b$ correlation
for the \Lya forest clouds.
We summarize our results in \S6.

All wavelengths in this paper are vacuum values, and all observed
wavelengths have been reduced to the solar rest frame. We use
$H_{0}$=100 \kms Mpc$^{-1}$ and $q_{0}$=0.5 where necessary.

\medskip
\centerline{2. SPECTROSCOPIC OBSERVATIONS}
\medskip

Here we present spectra of quasar HS~1946+7658 with intermediate resolution
\break
(FWHM $\sim$ 1.9~\AA\ -- 5.3~\AA), from 3240~\AA\ to 10570~\AA,  and high
resolution (FWHM $\sim$ 10 \kms, or $\sim$ 0.14--0.24~\AA) from 4220~\AA\
to 7251~\AA.

Table 1 is a journal of observations.  All spectra described
in this paper were obtained using the Lick 3-meter Shane telescope.
The intermediate resolution spectra
were obtained in 11.56 hours with the Cassegrain
{\it Kast Double Spectrograph} during a QSO identification program
described by Tytler \etal (1993), and the Coud\'e {\it Hamilton Echelle}
spectra were obtained in 11 hours during two nights, one in 1992,
and the other in 1993.

\vfill\eject
\medskip
\centerline{2.1 {\sl  Kast Spectra with 1.9 -- 5.3~\AA\ FWHM}}
\medskip

The superb new Kast Cassegrain double spectrograph records blue and red
spectra simultaneously. We used a dichroic beam splitter with a nominal
wavelength of 4600~\AA.  The red side (4000 -- 11000~\AA) light path is
identical to the former Lick Cassegrain UV Schmidt spectrograph
(Miller \& Stone 1987), with a silvered mirror collimator, reflection
gratings and a Schmidt camera. The blue side (3000 -- 7000~\AA\ )
has an identical (aluminized) collimator, transmission grisms, and a
refracting camera. Thinned Reticon 1200$\times$400 CCDs were used in
each of the cameras, with nine electrons readout noise on the red
side and seven on the blue.

In Table 1 we describe the five setups, A to E,
which we used to cover the wavelength range
from 3240 -- 10570~\AA. We used 1.5 arcseconds slit
for all Kast observations, and we rotated the spectrograph to
align the slit with the mid-exposure parallactic angle. The sky was
clear and the seeing ranged from 1.3 to 2.0 arcseconds with an average
1.6 arcseconds.


\medskip
\centerline{2.2 \sl Hamilton Echelle}
\medskip

The {\it Hamilton Echelle Spectrograph} is described by Vogt
(1987) and Misch (1991).
This instrument lives in the Coud\'e room, is fed by the Coud\'e
focus of the 3-m telescope.  An off-axis mirror
produces an 8.15-inch collimated beam which is dispersed by
an $8 \times 12$ inch 31.6~g/mm echelle which has a blaze angle of
$64.^o$7. The light is cross-dispersed by two UBK7 glass (UV absorbing)
prisms, and is focused by a folded Schmidt camera.
We used a thinned backside illuminated Texas Instruments 12~mm square
$800\times800$ CCD with $15~\mu$ pixels and seven electron readout noise.
Ignoring slit losses, the telescope -- instrument -- detector system
has a measured throughput which rises steeply in the blue, from
0.2\% at 3400~\AA\ , to 1\% at 3800~\AA\ and 4\% at 4200~\AA.
It has a flat peak of 5.7\% from 5000 -- 6200~\AA\ , and a gradual
linear decline to 2\% at 8200~\AA\  and 1\% at 9400~\AA\ (Misch 1991).

The CCD was binned $ 2 \times 2 $ pixels on the chip prior to readout.
A major advantage of an 8-inch beam on a 120-inch telescope is the ability
to get high resolution with a large slit. The Hamilton has a
(resolution) $\times$ (slit~width) ``throughput'' of RS=50,600 arcseconds,
which is slightly larger than UCLES on the AAT, and UES on the William
Herschel, and HIRES on the Keck I, because they all have larger primary
mirrors. But comparison between telescopes should actually be made using
the dimensionless number, (RS/seeing), which incorporates the advantages
of good seeing.

We took full advantage of the high throughput of this instrument
with a wide 2.5 arcseconds slit, which projected to two binned pixels
($2\times 2 \times 15~\mu = 60~\mu $) and gave an effective resolution
of about two binned pixels which is 0.16~\AA, or 10 \kms at
$\lambda \sim 4800$~\AA.

The exposures were centered on the meridian and the seeing was about
1.8 arcseconds. We did not use an image rotator or atmospheric
dispersion corrector, so some blue flux will have been lost at the
slit. The five mirror Coud\'e system was used because the quasar is
above the 52 degree declination limit of the three-mirror train.
We expect that the telescope throughput with the five-mirror system
is only about 60\% of that with the three-mirror train because of the
additional reflections and some misalignment between the three and
five-mirror systems (Misch, private communications).

In 1992 we took a single 4.97 hour exposure which covered
4220 -- 5260~\AA\ in echelle orders 135 -- 109. This had very low SNR
in the blue (SNR = 1 per 0.07~\AA\ pixel at $\lambda$=4255~\AA\ ),
but was good farther to the red (SNR=8 at 4930~\AA\ ).

In 1993 we took a 6 hour exposure of 4517 -- 7251~\AA\ (echelle orders
from 126 to 79). In the overlap region from 4517--5260~\AA\ , half of
which is in the \Lya forest, the SNR rises from 3 in the blue, to a peak
of 16 at 4900~\AA\ in the \Lya emission line.

The echelle spectra have gaps between all orders because the 12~mm
TI CCD is too narrow.
In Table 2 we list the measured wavelength range of each order, and the
SNR near the order center.

\medskip
\centerline{3. DATA REDUCTION}
\medskip

Data reduction was carried out with the NOAO IRAF software package.
We reduced the intermediate resolution spectra in the usual way.
We removed cosmic rays and divided by a flat field.  The spectra were
optimally extracted with a routine similar to that described by
Horne (1986). A 1~$\sigma$ uncertainty array was also obtained.
For wavelength calibration we used fifth
or higher order polynomial to fit spectra of the Ne, He, Ar and Hg
comparison lamps.  The typical residual in the fit was about
0.1 -- 0.2~\AA, depending on resolution of the spectrum.
We obtained an approximate flux scale by observing a spectrophotometric
standard star although the absolute level of the flux scale is
unreliable because we used a 1.5 arcseconds slit, and the standard star
is much brighter than, and at different air mass from,  the QSO. The
spectra  with overlapping  wavelength coverage were rebinned to a common
linear wavelength scale with about the same bin size as before rebinning,
weighted by their variance , and summed. The resulting spectra are shown
in Figure 1(a--e). The flux is $f_{\nu}$ in units of micro-Jansky, and
the wavelengths are vacuum heliocentric values. The data have not been
smoothed, and we show individual pixels.

We reduced the high resolution echelle spectra in the same way.
We extracted the spectral data using an optimal technique.
Each order was extracted from about 4 binned pixels along the slit,
while the background (sky plus dark current) was the
average over four orders of the 5--6 binned pixels in the minima
between orders.
A 1~$\sigma$ uncertainty array was also obtained.  We used Th-Ar
lamp spectra obtained before and after each QSO exposure to determine
wavelength scale.  Since each echelle order covers only about 30~\AA,
we used a linear fits which gave residuals of less than 0.01~\AA.

We obtain only one spectrum each night. They differ in wavelength coverage,
but there is substantial overlap in each of orders 126--109, where we
combined them because they had similar SNR.

In Figure 2 we show all 31 orders which show absorption lines. The other
26 orders which are all redward of the \Lya emission line are not shown.
All orders are divided by continua which are  cubic spline fits.
We ignored high sigma pixels, and fitted the splines to the
mean continuum in apparently line free regions,
but occasionally we had to add a point by eye to
jump over strong absorption lines.
The data have not been smoothed, and we show individual binned pixels.

\medskip
\centerline{3.1 {\sl Comments on the Spectra}}
\medskip

A Lyman Limit system  is clearly seen at \L $\sim$ 3520~\AA\ in
the setup-A spectrum (Figure 1a). In the setup-B spectrum most of
the absorption lines are \Lya forest. A strong absorption feature
at \L $\sim$ 4673~\AA\ is the DLA  line. The core of
this line should go to zero flux but it does not, because of bad
sky subtraction, which we were unable to correct.

Three percent CCD fringing can be seen from about 6600~\AA\
to 7100~\AA\ in the setup-D.  We did attempt to correct for
atmospheric absorption, especially the B band (6867~\AA), A
band (7600~\AA) and OH absorption at 7160--7340~\AA, but the B
and the A bands can still be seen in the setup-D spectrum at
\L=6875~\AA\
and \L=7608~\AA. The absorption lines (No. 10, 11 \& 12) in the setup
D spectrum are probably not real because of atmospheric OH absorption.
We did not correct the atmospheric absorption in the setup-E spectrum,
which is dominated by the OH absorption at wavelengths 9000--9900~\AA.


\medskip
\centerline{3.2 {\sl Emission Lines}}
\medskip

The setup-C spectrum
covers the major emission lines: \Lya, N~V, O~I and Si~IV + O~VI].
Other emission lines such as C~IV and C~III are seen in the setup-D
spectrum.
In Table 3 we give the peak positions and the equivalent widths, corrected
for absorptions, of the prominent emission lines.
All the emission lines have extended blue wings and steep
red wings, and none of the equivalent widths are well determined
because the lines are weak or blended. There is no obvious peak
for N~V at \L$\sim$5120~\AA, and \Lya has
a double peak. The errors on the measurements of the line peak positions
are typically 1--2~\AA. The errors on equivalent width are at least 10\%.
The ion laboratory wavelengths are adopted from Table 8 of
Tytler \& Fan (1992).
The final average emission redshift is mean of O~I, Si~IV+O~VI],
C~IV and C~III] lines weighted by their equivalent widths.
We do not use H~I, N~V or the He~II+[O~III] lines,
because they are blended. The mean emission redshift is
$\langle z_{em} \rangle$=3.051 .

\medskip
\centerline{3.3 {\sl Absorption Lines}}
\medskip

The absorption lines in the intermediate resolution spectra were analyzed
using the techniques described by Young \etal (1979) and Tytler (1982).
The continuum was fitted with a cubic spline, and was sometimes adjusted
to obtain a more precise fit, especially in emission line regions.
The absorption lines were selected and measured only if they exceeded
4.5$\sigma$(W), where $\sigma$(W) is the random
error in the measurement of the observed equivalent width W$_{obs}$.
In Table 4 we list the observed wavelengths, equivalent widths and their
1~$\sigma$ uncertainties, as well as the SNR of the continuum and the
line identifications.

We attempted to separate blended lines into their components.
The equivalent widths tabulated for these components cannot be
considered accurate, but the sum of the equivalent width should be
good. In addition the errors on the wavelengths are probably larger
than quoted for some partially blended lines

For the echelle spectral data we did not try to measure absorption line
centers and equivalent widths. Instead we used $\chi^{2}$ fits to
determine the line centers.
Since our spectral resolution allows us to identify the metal lines
in the \Lya forest, we assume that most unidentified lines are \Lya .

\medskip
\centerline{3.4 {\sl Errors in the Wavelength Scale}}
\medskip

Errors in the wavelength calibrations hinder the
identification of absorption lines, especially in the low
resolution spectra. Since the five intermediate resolution spectra
overlap, we checked the wavelength scales.

We found 24 absorption lines in common in setups A and B, and we
list the wavelengths for each in Table 4.
Their average wavelength difference is  $-0.07\pm0.14$~\AA, which is
about 0.2 pixels.

We also found 15 absorption lines in common in both setups B and C.
Since the setup-B spectrum has better resolution than the setup-C
spectrum, we list only the setup B absorption lines in Table 4, but
the average wavelength difference is  $1.05\pm0.14$~\AA.
This is about one pixel shift, which is not
unexpected because we are at the ends of the spectra.

To check our wavelength calibration in setup-E we measure atomic OH
sky emission line wavelengths in our extracted sky spectrum, and
compare them with the wavelengths given by Osterbrock \& Martel (1992).
The average wavelength difference is
$\langle\Delta\lambda\rangle = 0.07\pm0.04$~\AA, with $\sigma=0.07$~\AA.

\medskip
\centerline{3.5 {\sl Absorption Line Parameters}}
\medskip

We have measured absorption line parameters such as absorption redshift
($z_{abs}$),
\break
Doppler parameter ($b=\sqrt{2}\sigma$) and column density
($N$), using Voigt profile fits to the echelle spectra.
Each absorption line was fitted with a Voigt profile convolved with
the instrumental profile and multiplied by the continuum.
The line parameters $z_{abs}$, $b$ and $N$ were determined from the
minimum $\chi^{2}$. The fitting code includes the MINUIT
package from the CERN library which uses several methods to find the
global minimum. For further details on the minimization and error
analysis methods see James \& Roos (1975).

We estimated the continuum for each echelle order with
a cubic spline fit to spectral regions which were apparently line
free.  But in the wavelength region from 4665~\AA\ to 4700~\AA,
where the damping wing of a DLA line covers the whole order, the
continuum was estimated by comparing the line structure with the
intermediate resolution spectrum, and with the continua of the
neighboring orders. There is  nearly zero flux at $\lambda \sim 4673$~\AA\
in the core region of the DLA line.

In Table 5, we list the results of our fits to these absorption lines,
together with their cross identifications in the spectra from
setups B, C \& D.
These fits are also shown in Figure 2. We used the minimum number
of components which gave an acceptable fit.

\medskip
\centerline{4. HEAVY ELEMENT ABSORPTION SYSTEMS}
\medskip

First we discuss the individual absorption systems in \S4.1, then in \S4.2 we
estimate the metal abundance in the DLA  system at $z_{abs}$=2.8443 .

\medskip
\centerline{4.1 {\sl Comments on Individual Absorption Systems}}
\medskip

	Hagen \etal (1992) found three unequivocal metal absorption
systems at $z_{abs}$ = 1.738, 2.843, and 3.048. They claimed that the \Lya
absorption feature at $\sim$4670~\AA, which belongs to the 2.843 system,
splits into several components. Our spectra confirm these three systems
and we find  three new metal line systems at $z_{abs}$=1.1190,
2.6444, 2.8927, all with weak lines. We also see an LLS at
\L $\sim$ 3520~\AA\ which is part of the 2.843 system. Two of these
absorption systems have multi-component structure. We now discuss each
separately. Table 6 is a summary of the ion lines seen in each system.
We omit the many lines which we identified in Tables 4 and 5
in the \Lya forest.
Column densities and $b$ values are given for lines in the echelle spectra.

{\bf $z_{abs}$=1.1190}. --- We identify a Mg~II doublet (\L\L2796, 2803)
in the setup-C spectrum. The correspond strong Fe~II absorption lines
such as  Fe~II~\L2382, \L2600, \L2344 are marginally visible at the
expected positions, but they are too weak to be listed.  The Voigt
profile fit to the Mg~II(\L2803) in order 96 (line No. 152) indicates
that the line has $b$=11 \kms.  The Mg~II \L2796 line was not covered
in our echelle spectra.

{\bf $z_{abs}$=1.7382}. ---
This system is certain because
we identify all five strong Fe~II lines (\L2344, \L2374, \L2382,
\L2586 and \L2600), Mg~II (\L\L2796, 2803; note that
\L2796 is blended with the atmospheric A-band), and Al~III \L1854.
Blueward of the \Lya emission line we
identify C~IV (\L\L1548,1550), Si~IV (\L\L1393, 1402), Al~II~\L1670,
Fe~II~\L1608, Si~II~\L1526, Ni~II~\L1370, C~I~\L1656
and C~I~\L1329 lines.
The  C~IV doublet lines
are about half an \AA ngstrom off with respect
to Fe~II lines, which is consistent with the wavelength calibration
errors which we presented in \S3.4. Some of the above lines in the
\Lya forest region are blended with the lines of other systems.

This system may be damped, because we see weak Cr~II lines,
but the DLA line will never be seen because it
is below the LLS at \L $\sim$ 3520~\AA\ in the $z_{abs}=2.844$ complex.
In the intermediate resolution spectrum two Cr~II lines look real,
but their strengths are reversed:
\L2056.25 (f=0.140, W$_{obs}=0.14$\AA\ ), and \L 2066.16 (f=0.0698,
W$_{obs}=0.17$\AA\ ).
The third Cr line at \L2062.23 (f=0.105)
is lost in the blue wing of  line 11 in setup C.

In the echelle spectrum the Cr \L 2066 has the same W$_{obs}=0.17$\AA\ ,
but it is shallow and about 70 \kms wide,
sufficient that it may be two lines, which would explain why it has
a larger $W$ than Cr~II \L2056, which should be the stronger of the two.
Our best guess is that \L2066 has W$_{obs} \simeq$ 0.09 -- 0.14\AA .
Cr~II \L2056 is in an inter-order gap, and \L2062 has
W$_{obs}\simeq 0.06$\AA which
is weaker than expected. If the Cr~II lines are real, then the
 column density is about
$10^{13}$ \cm-2 with a range of about $(0.6 -2) \times 10^{13}$\cm-2.

Pettini \etal (1990b) found that the Cr abundance is about 0.01 of solar
in one DLA system with column density log $N$(H~I)=21.4,
at $z_{abs}$=2.3091, toward QSO PHL~957.  Since most gaseous Cr will be
Cr$^{+}$ in DLA systems, this means that gaseous Cr is underabundant by
a factor of 100.  The recent Pettini \etal (1994) study of 17
DLA systems with column densities of
(1--60) $\times 10^{20}$ \cm-2 shows that the PHL 957 DLA is typical,
and that the average DLA has an abundance of [Cr/H] $< -1.75$.
With the above Cr~II column density, and [Cr/H]=$-2$, the hydrogen column
density would be
$N$(H~I) $\sim$ 1.7$ \times 10^{21}$ cm$^{-2}$.
Therefore the system would be a DLA if the Cr lines are real.


Order 101 of our echelle spectra covers Cr~II(\L\L2062, 2066) lines,
but the SNR is not high enough for us to see the lines.

We see in the echelle spectra that the Fe~II (\L2344, \L2374 and \L2600)
lines all have simple velocity structure. Voigt profile fits to these
lines give $b$=16 \kms.  We estimate the gas phase abundance
[Fe$^+$/H] $\sim -2.6$.

{\bf $z_{abs}$=2.6444}. --- We identify only the C~IV doublet
(\L\L1548, 1550) absorption lines in this system.
They are lines 146 and 149 in echelle order 101,
although line 146 is slightly blended with a component
of the absorption complex at $z_{abs}$=3.05. In setup-C we only
identify line 12 as the C~IV(\L1550). The C~IV(\L1548) line is
embedded in the strong line 11. A fit to the line profile of the
C~IV doublet gives $b$=13 \kms. Line 27 in echelle order 129 is
the \Lya absorption line. We find [C$^{+++}$/H$^0$] $\sim -0.5$,
although the H~I column density is not very accurate because it
is blended with other \Lya lines. We can not determine the total
gas phase abundances because we did not know the ionization.

{\bf $z_{abs}$=2.844 4-component complex}. --- ($z_{abs}=2.843$ in
Hagen \etal 1992) This system contains the conspicuous DLA line at
\L $\sim$ 4673~\AA . The echelle spectrum shows the damping wing of
the line (4665~\AA\ -- 4700~\AA), and
the intermediate resolution spectra show the H Lyman series lines up
to Lyman-9~\L920. Metal ions
including C~IV (\L\L1548, 1550), Si~IV (\L\L1393, 1402), Si~II~\L1526,
C~II~\L1334 and O~I~\L1302 were all found redward of the \Lya emission
line in setup-C. Other metal species such as Si~II
(\L\L1190, 1193), O~VI (\L\L 1031, 1037), Si~II~\L1260, Si~III~\L1206,
N~II~\L1083 and C~III~\L977 were found in the \Lya forest region.

The echelle spectra cover most of the above lines, plus
Al~II(\L1670) and the fine-structure line C~II$^*$(\L1335).
These spectra also resolve most of the
strong lines in the intermediate resolution spectra.
In Figure 3, we show that the system contains at least four velocity
components at $z_{abs}$=2.8434, 2.8438, 2.8443 and 2.8452.
The $z_{abs}=2.8443$ components probably dominates the total column
density because it shows the strongest low ionization lines. In the
$z_{abs}=2.8434$ component the Si~II(\L1526) line is very weak (line
150 in order 97), and would not have been identified if we did not
know the redshift.

We find that the Lyman limit edge at \L $\sim$ 3520~\AA\ is
associated with this DLA complex. In Figure 4 we show a simultaneous
fit to the H Lyman series absorption lines at $z_{abs}=2.8443$ with
log $N$(H~I)=19.8 and $b_{\rm H}$=35 \kms. This shows that the LLS
is associated with this absorbing cloud.

Column densities, velocity dispersions and abundances in this system
are discussed in \S 4.2 .

{\bf $z_{abs}$=2.8927}. ---  This is a new system. In the intermediate
resolution spectra we identify a secure C~IV doublet (\L\L1548, 1550),
and in the \Lya forest region, the Si~III~\L1206 and C~III~\L977 lines.
We also see the H Lyman series lines up to Lyman-10~\L919.
The \Lya absorption line may blend with \Lya lines of two other
systems in the echelle spectra.
The C~IV~\L1548 line is apparently broad in echelle
order 95, which we do not show because the SNR is very low near
the line.

{\bf $z_{abs}$=3.050 3-component complex}. --- In the intermediate
resolution spectra we identify many metal species including
C~IV (\L\L1548, 1550), Si~IV (\L\L1393, 1402), N~V (\L\L1238, 1242),
C~II~\L1334 and Si~II~\L1260 redward of \Lya emission line, and
Si~III~\L1206,
O~VI (\L\L1031, 1036), N~II~\L1083, N~III~\L989, Si~II~\L989
and C~III~\L977 in the \Lya forest region. We also identify three
strong lines: \Lya, \Lyb and \Lyg.

The \Lya absorption line (No. 97 in Fig. 1b) is very strong compared
to other \Lya forest lines. This suggests complex velocity structure,
which we confirm in the echelle spectra, where we see three components
at $z_{abs}$=3.0483, 3.0495 and 3.0504 .

In order 112 two Si~II~\L1260 lines at 5104~\AA\ and 5105~\AA\ give
redshifts
$z_{abs}$=3.0495 and 3.0503.
They appear as a single strong line in setup-C (Figure 1c, line No. 5).
The N~V~\L1238 lines in these components are seen in order 114, although
the $z_{abs}$=3.0504 component is relative weak. The third component at
$z_{abs}$=3.0483 is seen from the C~IV doublet in order 91, and also from
Si~IV(\L1393) in order 101, but the latter may be blended with C~IV(\L1548)
at $z_{abs}$=2.6444.

The \Lya absorption line of the system at \L=4923~\AA\ (echelle order 116)
is clearly saturated or a blend. We suggest that there are at least two
components because it is slightly asymmetric. In Figure 5 we show the
profiles of several metal lines in this absorption complex.

\bigskip
In Figure 1e we show the expected positions of the five strongest
Fe~II lines (\L2344, \L2374, \L2382, \L2586 and \L2600) in the four
absorption systems with $z_{abs} > 2.5$. Although we know that our
wavelength scale is good (\S 3.4), we do not see any of these lines,
which is reasonable for the weak line systems at $z_{abs}$=2.6444
and 2.8927, but for the absorption complexes at $z_{abs}=2.844$ and
$z_{abs}=3.050$ it is a surprise because we see Si~II lines. Perhaps
iron is underabundant.

\vfill\eject
\medskip
\centerline{4.2 {\sl Metal Abundance of the $z_{abs}$=2.8443} }
\medskip

In Table 7 we give column densities and $b$ values obtained by Voigt
profile fits to the metal lines in this damped system.

\medskip
\centerline{4.2.1 {\sl H~I Column Density}}
\medskip

We can place four limits on the H~I column density $N$(H~I).

1. The Lyman continuum is completely absorbed  to the blue of
\L$\sim$3520~\AA\ in Figure 1a, which indicates that the optical depth is
$\tau_{\rm LL} > 3.0$, and the neutral hydrogen column density is
$N$(H~I) $>$ 5$\times 10^{17}$ cm$^{-2}$.

2. The equivalent width (W$_{obs}$=21.67~\AA) of the \Lya absorption
line in the intermediate resolution spectrum corresponds to
log $N$(H~I) $\sim$ 19.8  cm$^{-2}$ for a damped line.

3. A Voigt profile fit to the red wing of the line in the echelle
spectrum gives log $N$(H~I) $\simeq$ 20.2$\pm$0.4 cm$^{-2}$.  We included
components for the three adjacent \Lya forest lines (lines 80, 81, 82
in figure 2) in this fit, and we allowed the absorption redshift, the
H~I column density and $b$ to vary for each component.

4. In Figure 4 we show that $b_{\rm H}$=35 \kms and log $N$(H~I)=19.8
give a good fit to H Lyman series lines 5 -- 9. Most of the H Lyman series
lines in this system may blend with \Lya lines of other systems,
therefore the actual $b_{\rm H}$ may be less than the value of 35 \kms,
which we list in Table 7.

In Figure 6 we show curve of growth with log $N$(H~I)=19.8 cm$^{-2}$ and
$b$=35, 41 \kms respectively.  The line equivalent widths are all from
the intermediate resolution spectra, so the components of the system
are not resolved.  The {\it stars} represent the H Lyman series lines,
from right to left, \Lya, \Lyb, \Lyg, \Lyd etc. The \Lyg and \Lyd lines
are obvious blended, while the higher order lines may also be blended.
The best value of $b_{\rm H}=35$ \kms for the H Lyman series is consistent with
that of $b=41$ \kms obtained from the curve of growth fit to the metal
lines.

We choose log $N$(H~I)=20.2 cm$^{-2}$ for the following abundance estimates,
because the value of $N$(H~I)$\simeq 19.8$
from the intermediate resolution spectra is less accurate:
the DLA line does not have zero flux at its center, so the $W_r$ is not
reliable, and the fit to the Lyman series is not very sensitive to $N$(H~I).

\medskip
\centerline{4.2.2 {\sl Column Densities of Metals}}
\medskip

We saw in \S4.1 that the $z_{abs}$=2.844 DLA complex has four
components spanning 150 \kms.  The lines of the low ionization ions
O~I, C~II, Si~II and Al~II are concentrated in the component at
$z_{abs}$=2.8443, which has an average $b=10\pm1$ \kms,
while the high ionization
ions C~IV and Si~IV are found mostly in the three other components.
The absorption redshift of the DLA line is consistent with the low
ionization component. This kind of dual structure is typical of DLA
systems (Turnshek \etal 1989, Lu \etal 1993, Wolfe \etal 1993) which
usually have both quiescent and  turbulent components. These papers
report that the quiescent component contains most of the H~I, and
produces most low ionization lines with a typical Doppler parameter
$b \sim 15$ \kms, while the turbulent component contains a small
fraction  of the H~I, has complex velocity structure and has both high
and low ionization absorption lines with a typical $b > 30$ \kms.

We will assume that the velocity component at $z_{abs}$=2.8443
contains most of the H~I, and we will estimate metal abundances
for this component alone.  In the top of Table 7 we list the metal
lines identified in this velocity component.  We cannot estimate
abundances for the other components because we do not know how
much neutral or ionized Hydrogen they contain.

We identify four Si~II lines (\L1190, \L1193, \L1304 and \L1526),
but the strongest
\break
Si~II~(\L1260) line is blended with \Lya forest
lines 107 and 108 in echelle order 118.  The four Si~II lines
should give fairly accurate column densities because  they have
simple profiles, but they actually cover an order of magnitude.
These differences are
not significant, because all are within $1 \sigma$ of
the weighted mean and the $\chi^2_3 $ value of 4.11 would be
exceeded 25\% of the time. Therefore we
use the weighted mean of log $N$(Si~II)=13.2$\pm$0.3.

The only Si~III line is \L 1206. We tried to fit it with four blended
components, but the $b=38$ \kms for the 2.8443 component is large
compared with the mean $b=10$ \kms for the other lines in this component.
Perhaps the Si~III is mostly in the more ionized gas which gave the
value of $b$=35 \kms that we obtained from the curve of growth fit to
the H Lyman series lines.

\medskip
\centerline{4.2.3 {\sl Curve of Growth Check of Column Densities}}
\medskip

We have also used the intermediate resolution spectra to check the column
densities determined from the echelle spectra. In the lower half of Table 7
we give
results from curve of growth fits to all the metal lines redward of the
\Lya emission line in this DLA system.  These column densities apply to
all four components of the system.  The single Doppler parameter is
$b=41\pm5$ \kms, which is substantially larger than the $b=10$ \kms of
the $z_{abs}$=2.8443 component, but is much less than the 150 \kms covered
by all four components. This is because the curve of growth is determined
by the C~IV and Si~IV doublets
which have FWHM $\sim$ 40 \kms from the two strong
component at $z_{abs}=2.8434, 2.8438$. This $b$ value is close to the
$b_{\rm H}=35\pm5$ \kms obtained
from Voigt profile fits to the higher order H Lyman series lines 5--9 in the
intermediate resolution spectra.

We find that the curve of growth column densities of O~I, C~II and Si~II
are all very close to the values which we obtained from the echelle data
for the 2.8443 component.
Jenkins (1986) showed that if there is one main component,
which is not strongly saturated, which is the case here, then the
curve of growth column density should be good.
We show the curve of growth in Figure 6.

\medskip
\centerline{4.2.4 {\sl C~II$^*$ and the Gas Density}}
\medskip

In \S4.2.5 we will use the photoionization code CLOUDY (Ferland 1991)
to determine how the abundances of the ions depend on ionization parameter
$\Gamma$.  The hydrogen number density $n_{\rm H}$ is needed to estimate
these parameters.
The gas density in the cloud can be derived from the ratio of
$N$(C~II$^*$)/$N$(C~II) (Bahcall \& Wolf 1968).
The existence of fine-structure line
C~II$^*$(\L1335) in this DLA system indicates that either
(1) the gas density is large enough for collisional excitation of the
fine-structure level, or
(2) the ambient ionizing flux populates the  fine-structure state
(Bahcall \& Wolf 1968). The C~II$^*$ fine-structure level can be
populated by other processes such as direct pumping by IR photons and
recombination of C~III (Morris \etal 1986), but since this system is
a DLA system which has large H~I column density, we consider only
collisional excitation.  Using the expression given by Morris \etal,
$$
{N({\rm C~II^*}) \over N({\rm C~II})} = 3.9\times 10^{-2}n_e[1+(0.22n_p/n_e)],
$$
and for $n_e \approx n_p \approx n_{\rm H}$ we get,
$$
n_{\rm H} =21\left({N({\rm C~II^*)} \over N({\rm C~II})}\right).
$$
The total column densities of C~II and C~II$^*$ give a hydrogen number
density of $n_{\rm H} \sim$ 3.8 cm$^{-3}$. We regard the column density of
$N$(C~II$^*$) obtained from the echelle spectra as an upper
limit because the line may be saturated or it may blended with other
components.

The C~II$^*$ line seen in the echelle data is very weak, and it is
offset in velocity relative to O~I, C~II, Si~II and Al~II, so we
wonder if it is real. The  line is not visible in the intermediate
resolution spectrum, with a 3~$\sigma$ limit of W$_{r}$=24~m\AA\
which corresponds to $N$(C~II$^*$)=$1.3 \times 10^{13}$ cm$^{-2}$
(assume it is on the linear part of curve of growth), but  this is
consistent with the detection in the echelle spectrum, and we are
confident that the fine-structure line is real.

The line does not belong to any of the five other metal systems,
and it is probably not C~IV~\L1548 or  Mg~II~\L2796 because the doublet
pairs are missing, but it could be
C~IV~\L1550 or  Mg~II~\L2803, because the doublet pair would then be in the
inter-order gap.

The redshift of the C~II$^*$ line with respect to C~II(\L1334) line is small,
only about $\sim$ 15 \kms, but it is three pixels and it is real
because both lines are in the same echelle order and we see no problems
with the wavelength scale, which is linear.

In Figure 7 we show the C~II and C~II$^*$ line profiles and the ratio
$$
R(v) = {{N_{{\rm CII^*}}(v) \Delta v} \over {N_{{\rm CII}}(v) \Delta v} },
$$
as a function of velocity with respect to $z=2.8443$, which is near the center
of the C~II line. Here $N(v)dv=N \phi(v)dv$ and $\int \phi (v) dv=1$. The
velocity shift indicates that the density or excitation temperature
is higher at higher velocities.
To calculate $R(v)$ and the temperature or density gradients as a
function of velocity we should first subtract the thermal line
broadening, but we prefer to wait for Keck data with much higher
SNR. Here we simply note that if the range is $0.1 \le R(v) \le 0.4$,
then the range of excitation temperature would be
$32 \le T_{ex} \le 100$~K, the density would be
$2.1 \le n_{\rm H} \le 8.4$~cm$^{-3}$, and the DLA clouds would be
6 -- 25~pc thick, reasonable for a single cloud in a cold spiral galaxy disk.
(The disk scale height is not the expected thickness,
because the quasar light beam is too narrow.)

\medskip
\centerline{4.2.5 {\sl Level of Ionization}}
\medskip

We used Ferland's photoionization code CLOUDY to calculate the expected
column density of different ions as a function of photoionization parameter
$\Gamma \equiv n_{\gamma}/n_{\rm H}$. We used the Bechtold \etal
(1987) ``medium'' spectrum for the incident radiation field (also
see Steidel 1990), and a grid search to find the parameters which
match our observed column density ratios. The column
densities of $N$(Si~II), $N$(C~II) and $N$(O~I) do not change much
for a wide range of $\Gamma$ because they are tied to $N$(H~I),
which we know. Their ratios do not give a good
estimate of $\Gamma$, although they do give good metal abundance
estimates. But $N$(Al~II), $N$(C~IV) and $N$(Si~IV) depend on the
ionization strongly, and their column density ratios should give
a good estimate of the ionization.

We will use nine column density ratios selected from ions
C~II, C~IV, O~I, Si~II, Si~IV
and Al~II to estimate the ionization parameter in the DLA system at
$z_{abs}=2.8443$. We note in Figure 3 that the C~IV and Si~IV lines
at this velocity are not very significant, so we cannot fit Voigt
profiles to get their column densities. Instead we measure 5 pixel
(2$\times$FWHM+1) equivalent widths and estimate column
densities of log $N$(C~IV) $\leq 13$ and log $N$(Si~IV) $\leq 12.6$,
which we list in Table 7. These column densities are lower limits because we
use the linear part of the curve of growth; but they would be an upper limits
if most of the apparent absorption were noise.

We searched a grid region of log $N$(H~I)=$20.2\pm0.4$ and
[M/H]=$-2.6\pm0.3$ (from  the next section) with a total hydrogen number
density of $n_{\rm H}=2.45$ cm$^{-3}$, the value which we estimated
at $z_{abs}=2.8443$ (see figure 7). We found a match to the column
density ratios for log $N$(H~I)=20.2, [M/H]=$-2.8$ and
$\Gamma \simeq -2.75$ with a range of ($-$3.1, $-$2.5). In Table 8 we list
these nine observed column density ratios, and  the ionization parameter
needed to match to the observed ratios.
We also list the expected column density ratios at $\Gamma=-2.75$.
In Figure 8 we show the expected column densities for four elements
-- C, O, Si and Al -- for log $N$(H~I)=20.2 and [M/H]=$-2.8$. We also
show our measurements of the column densities of O~I, C~II, S~II, Al~II,
C~IV and Si~IV.
We see that at $\Gamma=-2.75$ C~II is about ten times more abundant
than C~IV, and Si~II is about 4.5 times as abundant as Si~IV,
in agreement with the observations. But we also see that the C~II and
O~I are not the most abundant ions because  C~III (not covered by the
echelle spectrum)
is more abundant than C~II and O~III (not covered by either spectrum)
is more abundant
than O~I.

\medskip
\centerline{4.2.6 {\sl The Metal Abundances in the DLA is Low}}
\medskip

The column densities of absorption lines give gas phase abundances.
In the top panel of Table 7 we see that the low ionization absorption
lines dominate in this DLA system. If we assume that the $N$(H~I)
represents the total hydrogen column density, and that the column
densities in Table 7 represent the total column densities of their
elements, then we obtain gas phase abundances for O, C, Si and Al of
[O/H]  $\simeq -2.8\pm0.6$,
[C/H]  $\simeq -2.7\pm0.6$,
[Si/H] $\simeq -2.5\pm0.5$ and
[Al/H] $\simeq -2.5\pm0.5$.
Together they imply a gas phase metal abundance of
[M/H]  $\simeq -2.6\pm0.3$, which is low for a DLA system compared with
results of Pettini \etal (1994).
In the above estimation, the total column density of Carbon includes C~II,
C~II$^*$ and C~IV, and total column density of Silicon includes Si~II and
Si~IV,
but we consider only observed ions. If instead we use the results from the
ionization model to include ions which are not seen (e.g. H~II, C~III and
O~III) we obtain
[O/H]  $\simeq-2.7$,
[C/H]  $\simeq-2.9$,
[Si/H] $\simeq-2.7$ and
[Al/H] $\simeq-2.8$,
which are close to the abundances from observed ions alone, in part
because the omission of unobserved metal ions was compensated by the
omission of unobserved H~II.

Is the abundance really this low? Perhaps we failed to resolve
very low $b$ velocity components, or perhaps the metals are depleted
onto dust. Narrow components could not be seen in our data because
the SNR is low and the line centers are black for C~II~\L1334 and
O~I~\L1302 (Figure 3). The line center optical depth of these two
lines are $\tau_{0}=2.12$ and 2.54 respectively (Table 7) -- just
sufficient that a significant amount of gas could be missed in
unresolved narrow components.

But we do believe that very little gas is locking in dust grains, because
the abundances ratios are all close to cosmic.  In Table 9 we list
abundance ratios for C, O, Si and Al, together with typical ratios
for the ISM, where 99\% of Cr is in dust grains.
Our ratios are close to cosmic, and far from the depleted ISM values.
Aluminum provides the best test because it depletes strongly like
Cr in the ISM. We measure [Al/C]$=0.23$, [Al/O]$=0.29$ and [Al/Si]$=0.00$,
which are all consistent with no depletion.
In the ISM [Al/X] $< -0.5$ (Spitzer 1978; Jenkins 1987).
In this DLA system we do not see the enhancement of O with respect to
C claimed by Vogel \& Reimers (1993); we measure [O/C]=--0.06 for
observed ions and [O/C]=0.2 after ionization correction.

\medskip
\centerline{5. \Lya FOREST ABSORPTION SYSTEMS}
\medskip

In this section we will discuss the column density, velocity dispersion
distribution, and $N-b$ correlation of the \Lya forest lines.

\medskip
\centerline{5.1 \sl Number Density of \Lya Forest Lines}
\medskip

It is well established that the distribution of \Lya forest absorption
lines is roughly proportional to $(1+z)^\gamma$, with $\gamma$ between
2 and 3 for rest equivalent widths W$_{r} \geq 0.36$~\AA\ (Murdoch \etal
1986, Tytler 1987a, Bajtlik \etal 1988, Fan \& Tytler 1989, Lu \etal 1991,
Bechtold 1993). In our intermediate resolution spectra we find 25 \Lya
absorption lines with W$_{r} \geq 0.36$~\AA\ between the Ly$\beta$ and
the \Lya emission lines, excluding the DLA line region
$z_{abs}$=2.825 $-$ 2.866.

\vfill\eject
\medskip
\centerline{5.1.1 {\sl Proximity Effect}}
\medskip

Lines which have $z_{abs} \simeq z_{em}$ should be more highly ionized
on average, which is known as the proximity effect. The critical
luminosity distance $r_L$ to a QSO is defined by (see eqs. 4, 6, 7 of
Bajtlik \etal 1988):
$$
r_L = 2.58\left({L_{30} \over J_{-21}}\right)^{1 \over 2}h^{-1}_{100}{\rm Mpc},
$$
where $L_{30}$ is $L_{\nu}$ --  the intrinsic QSO luminosity at the
Lyman limit in unit of 10$^{30}$ (ergs s$^{-1}$ Hz$^{-1}$), and
$J_{-21}$ is $J_{\nu}$ -- the background radiation flux at the Lyman
Limit in unit of 10$^{-21}$ (ergs cm$^{-2}$ s$^{-1}$ Hz$^{-1}$ sr$^{-1}$).
This $r_L$ is the distance at which the local ionizing flux from the QSO
is at least 10\% of that from the IGM flux, or log $\omega \geq -1$ in
terms of the notation of Bajtlik \etal \Lya forest lines which are
closer than $r_L$ to the QSO should be more highly ionized, and hence
they should show less H~I.

For HS~1946+7658 we estimated that the continuum flux at 912~\AA\ of
QSO rest frame (\L$_{obs} \sim$3695~\AA)
is about 1.45$\times 10^{-26}$ ergs cm$^{-2}$ s$^{-1}$ Hz$^{-1}$ by
extrapolating the continuum from redward of the \Lya emission
line, which is consistent with the general $V$ versus $L_{\nu}(912)$
relationship of Tytler (1987a) for V=15.85 (Hagen \etal 1992) and
$z_{em}$=3.051. The proximity effect is then expected to extend to
about 20~$h_{100}^{-1}$~Mpc.

We see six \Lya lines in the intermediate resolution spectrum
within this 20$h_{100}^{-1}$ Mpc region, which is consistent with
expectations from the previous cited data.

\medskip
\centerline{5.1.2 {\sl \Lya Line Density Away From the QSO}}
\medskip

Beyond $r_L$ we expect 34 lines and see only 19, which is a 2.6$\sigma$
deficiency.  These lines lie in a redshift interval of $\Delta z=0.533$
(which excludes the DLA line region) from the range
$z_{abs}$=2.418 -- 2.992. The line density is $d{\cal N}/dz$=36$\pm$8
at the mean absorption redshift of $\langle z \rangle$=2.676. We obtained
the expected number of \Lya lines from $\gamma=2.75$ and $N_0=1.74$
of Lu \etal (1991), but $\gamma=1.85$ and $N_0=5.4$ from Bechtold (1993)
give a similar result.  We note that the line density is particularly
low in the vicinity of the DLA system, and we wonder if this might be
a general trend.

In Figure 9 we plot the \Lya line (W$_{r} \geq 0.36$~\AA)
distribution, in co-evolving coordinate $\chi_{\gamma}$, for
$\gamma = 2.7$, for 16 QSOs which all have DLA systems. They are ranked by
their emission redshifts which are from $z_{em}=2.08$ to 4.11. The parts of
the spectra which are obliterated by the DLA systems are indicated by the
horizontal bars.  The mean line density does appear to vary a lot on
scales of $\chi_{\gamma} \simeq 5$,
although previous studies have shown that there are no
significant variations on scales larger than 5,000 \kms
(Carswell \& Rees 1987; Ostriker, Bajtlik \& Duncan 1988; Babul 1991;
Webb \& Barcons 1991). We see that the
lack of lines in HS~1946+7658 is most pronounced at high redshifts,
as if the proximity effect extends far farther than we had thought, but we
do not see a lack of \Lya lines around other DLA systems.
For now we consider that the lack of lines is a statistical
fluctuation, but we suspect that there are real variations in the
line density.

\medskip
\centerline{5.2 {\sl Doppler Parameters and Column Densities in
The \Lya Forest}}
\medskip

There are 118 \Lya absorption lines in our echelle spectra, five of
which are associated with metal systems. (Here we include lines
influenced by the proximity effect). Of the 113 \Lya lines which
do not show metal lines, 17 are sufficiently saturated or blended that
a single Voigt profile fit does not give accurate column densities or
velocity dispersions, but the other 96 appear unsaturated, by which we
mean the line flux does not go to zero.

In Figure 10 we show the distribution of $b$ and
log $N$(H~I) for the \Lya lines. The unsaturated lines are shown as
{\it filled circles} if SNR$<$8 (\L $<$ 4700~\AA) and {\it crosses}
if SNR$>$8 (\L $>$ 4700~\AA). The saturated/blended lines are shown
as larger {\it open circles}, with an enclosed \lq+\rq\ if SNR$>$8.
We also show curves of constant rest equivalent widths,
W$_r$=0.40~\AA, 0.32~\AA, 0.20~\AA, 0.10~\AA\ and 0.05~\AA.
Note that only 9\% of the lines have $W_r \geq 0.4$\AA\ and that most have
$0.1 \leq W_r \leq 0.32$\AA, so our lines are much weaker than those which
comprise most intermediate resolution samples with $W_r \geq 0.36$\AA.

\medskip
\centerline{5.2.1 {\sl Proportion of low b lines}}
\medskip

\Lya absorption lines with $b \leq 20$ \kms are especially interesting
because they come from clouds which are cool and have low internal
motions. PHSM claimed that such lines were more common than had
previously been believed. In Table 10 we summarize our data and
other data with similar or better resolution from four other QSOs.
We list the fraction of \Lya lines with $b \leq 20$ \kms separately
for log $N$(H~I) above and below 13.5.

Our mean $b$ is in excellent agreement with the published values.
When we include the saturated lines, and those in low SNR spectra,
43\% of our lines with log $N$(H~I) $\leq$ 13.5  have $b \leq 20$ \kms.
This is intermediate between the values of 22\%, 29\%, 50\% and 53\%
of the four other QSOs, and the large QSO-to-QSO variation is easily
understood because log $N$(H~I)=13.5 is near the detection limit for
$b > 20$ \kms, so a slight increase in the SNR will yield many more
lines at low $N$.

But at log $N$(H~I) $>$ 13.5, 33\% of our lines have $b <$ 20 \kms,
which is much higher than the values of 5\%, 6\%, 10\% and 20\% found
in the four other QSOs. We believe this is because our SNR is very
low.  At $4700 \leq \lambda \leq 4950$~\AA\ where
$8 \leq {\rm SNR} \leq 16$
we have 43 \Lya lines, and only 14\% of these 29 lines with log
$N$(H~I) $>$ 13.5 have $b \leq 20$, which is consistent with published
values.

When SNR $\geq$ 8, we believe that the $b$ values from the Voigt profile
fits are usually well determined, but only 2 out of 7 unsaturated lines
with $b < 20$ \kms (numbers 91 \& 103)  are believed to have reliable $b$
values. The other 5 may be underestimated because either they are blended
with other lines or their line profiles are not well defined.

Our data do have sufficient SNR to find low $b$ lines, but not to avoid
bias in the measurements of $b$ and $N$(H~I) values.  Lines with
log $N$(H~I) $ > 13.5$ and $10 < b < 20$ \kms all have W$_r \geq 0.1$~\AA.
For \Lya lines the 1~$\sigma$ detection limit is
$$
\sigma({\rm W}_r) \simeq {{1215.67 \times {\rm FWHM}}
\over {\lambda_{obs} \times {\rm SNR}}}.
$$
To detect lines with a 4~$\sigma$ limit with FWHM=0.2~\AA\ we need
SNR $\geq$ 2.3 to get $\sigma({\rm W}_r) \leq 0.025$~\AA\ at 4300~\AA.
In all cases our spectra do have the SNR required to detect the listed
\Lya lines at 4$\sigma$ limit and most of these lines are also visible
in the very high SNR intermediate resolution spectra. But neither of
these conditions prevent the observed bias to low $b$ values. Only much
higher SNR can do that.

\medskip
\centerline{5.2.2 {\sl $N - b$ Correlation}}
\medskip

For the unsaturated \Lya lines ({\it solid dots and crosses}) we see a
correlation
between $b$ and $N$(H~I) with a correlation coefficient of 0.466, which
has a chance probability of $1.7 \times 10^{-5}$. This correlation is also
seen in the unsaturated \Lya lines of Q2126$-$158 (Giallongo \etal
1993, their figure 4). But for all \Lya lines ({\it all points})
the correlation coefficient shrinks to only $-0.02$, which is not
significant because it is expected in 86.3\% of uncorrelated data
sets. The $N$(H~I) and $b$ values clearly remain strong correlated, but
not in the simple linear way which is required to give a non-zero
correlation coefficient: a zero correlation coefficient does not mean
there is no correlation. The correlation coefficient vanishes because
the saturated lines have low $b$ for their large $N$, which is especially
true for the four lines with the largest $N$ in the low SNR data
(the {\it large empty circles}). We believe that we have overestimated the
column densities of these lines.

Rauch \etal (1993) used numerical simulation of the PHSM Q2206$-$199N
spectrum to show that the apparent $N - b$ correlation is an artifact
of the low SNR (see their figures 13, 14) which does not require an
intrinsic correlation. In their analysis the \Lya lines tend to move to
lower $b$ and $N$(H~I), or to larger $b$ and $N$(H~I).

In Figure 10, we can see that none of the unsaturated lines with
$b < 15$ \kms are from high SNR data, and all but one of the saturated
lines with
$b < 20$ \kms are from low SNR data. For the 36 high SNR unsaturated
lines ({\it crosses}) the $N - b$ correlation coefficient is 0.427, which
has a chance probability of $9.3\times 10^{-3}$, but if we remove the
outlier at $b=80$ \kms (line No. 85 in order 121) the correlation coefficient
drops to only  0.316 with a chance probability of 6.4\%, and if we keep the
outlier and also add the 7 saturated lines ({\it pluses}) from the high
SNR data the correlation coefficient drops farther to
0.169, with a 17\% chance probability for an uncorrelated data set.

\bigskip
We conclude that our spectra are not well suited to a study of
the $N - b$ correlation.  To study this, and the low $b$ \Lya lines,
we need higher resolution and much higher SNR.

\medskip
\centerline{6. SUMMARY}
\medskip

We have presented high (FWHM $\sim$ 10 \kms) and intermediate
(FWHM $\sim$ 1.9 --5.3~\AA) resolution spectra of QSO HS 1946+7658
from the Lick 3-m Shane telescope. Our spectra have amongst the best
high resolution coverage of the metal lines to the red of the \Lya
emission line.

We identified six metal and 113 \Lya absorption systems.
The metal absorption
complexes at $z_{abs}$=2.844 and 3.050 have multi-velocity
structure with 4 and 3 sub-components respectively, which span a total
velocity intervals of less than 200 \kms. The absorption component at
$z_{abs}=2.8443$ is a DLA system with H~I column density
log $N$(H~I)=$20.2\pm0.4$ cm$^{-2}$, an average $b=10\pm1$ \kms
and [M/H]$\simeq-2.8$.
We used a photoionization model to obtain this abundance, since
C~III and O~III are more abundant than C~II and O~I which we observed.
The abundance ratios for
C, O, Al and Si are all within a factor of two of solar, which
suggests that there is not much depletion onto dust grains.
We do not see an excess of O with respect to C, as was
reported in a few absorption systems (Vogel \& Reimers 1993).

In the DLA system, we find that the fine-structure
C~II$^*$(\L1335) line is redshifted about 15 \kms with respect to
the C~II(\L1334) line, which indicates that there is a
density or excitation temperature gradient within the absorbing cloud.
We estimate the total hydrogen density $2.1 < n_{\rm H} < 8.4$ cm$^{-3}$
from which we estimate that the thickness of this damped \Lya clouds
is about 6--25~pc, which is reasonable for a single cloud in a spiral galaxy
disk.

Our \Lya forest line sample shows a considerable fraction of narrow
lines with $b \leq 20$ \kms. But most of these narrow lines are artifacts of
our low SNR data. For the high SNR data we see the same number of
narrow lines as do others with similar data.
We agree with Rauch \etal (1993) that much higher SNR data is needed.

We see a correlation between column density $N$(H~I) and Doppler parameter
$b$ for our unsaturated \Lya forest absorption line sample.
But with the high SNR \Lya forest line sample, this correlation is
less impressive and we cannot tell if it is real or an artifact
of the low SNR.

\bigskip\medskip

We wish to thank James Burrous, Wayne Earthman and Keith Baker
at Lick Observatory for their helping during the observations.
We thank Anthony Misch for advice on echelle spectrograph
instrument. We thank G. Ferland for providing a copy of his CLOUDY
photoionization code, and we are very grateful to A. M. Wolfe for many
helpful discussions. We also thank Simon Morris for useful comments and
careful reading the manuscript. This work was supported in part by
NASA grant NAGW--2119.

\vfill\eject
%
%
\sub
\baselineskip=14pt
\pageno=24
$$\vbox{
\halign{
\tabskip=0.5em
\hskip 0.5em #\hfil & \hfil#\hfil &      #\hfil & #\hfil
      & \hfil#\hfil & \hfil#\hfil & \hfil#\hfil \cr
\multispan{7}\hfil {TABLE 1}\hfil\cr
\multispan{7}\hfil OBSERVATIONS \hfil\cr\cr
\noalign{ \hrule height .08em \vskip 2pt\hrule height .08em \vskip 6pt}
\omit\hfil Date \hfil & Exposure & \omit\hfil Setup \hfil & \omit\hfil
$\lambda$ range \hfil
                      & Slit  &\AA/pixel& FWHM~$^a$ \cr
\omit\hfil (UT)\hfil &   (s)    & & \omit\hfil (\AA)\hfil & width && (\AA)
\cr
\noalign{\vskip 6pt \hrule height .08em \vskip 6pt}
1992~Apr~24 & 3600 & A: 830/3880/b~$^b$ & 3241--4440    & 1.5$^{\prime\prime}$
& 1.1 & 2.5 \cr
1992~May~01 & 6000 & A: 830/3880/b     & 3241--4440     & 1.5$^{\prime\prime}$
& 1.1 & 2.5 \cr
1992~May~29 & 4000 & A: 830/3880/b     & 3241--4440     & 1.5$^{\prime\prime}$
& 1.1 & 2.5 \cr
1992~Oct~24 & 3600 & A: 830/3880/b     & 3241--4440     & 1.5$^{\prime\prime}$
& 1.1 & 2.5 \cr
1992~Jun~28 & 7200 & B: 831/8460/r~$^c$ & 4045--4963    & 1.5$^{\prime\prime}$
& 0.84& 1.9 \cr
1992~Apr~24 & 3600 & C: 1200/5000/r    & 4817--6176     & 1.5$^{\prime\prime}$
& 1.1 & 2.5 \cr
1992~May~01 & 6000 & C: 1200/5000/r    & 4817--6176     & 1.5$^{\prime\prime}$
& 1.1 & 2.5 \cr
1992~Oct~24 & 3600 & D: 600/7500/r     & 5990--8798     & 1.5$^{\prime\prime}$
& 2.3 & 5.3 \cr
1992~May~29 & 4000 & E: 831/8460/r     & 8610--10570    & 1.5$^{\prime\prime}$
& 1.65& 3.8 \cr
1992~Aug~05 &17900 & see Table 2 & 4220~$^d$--5261~$^e$ & 2.5$^{\prime\prime}$
& 0.08~$^f$& 0.14--0.18\cr
1993~Jun~29 &21600 & see Table 2 & 4517--7251~$^e$ & 2.5$^{\prime\prime}$ &
0.10$^f$& 0.15--0.24\cr
\noalign{\vskip 6pt \hrule height .08em \vskip 6pt}
\noalign{\hbox{$^a$ FWHM is typically 2.3 pixels for intermediate resolution
spectra and 2 pixels for} \vskip 6pt}
\noalign{\hbox{echelle spectra.} \vskip 6pt}
\noalign{\hbox{$^b$ This means ``the 830~g/mm grating, blazed at 3800~\AA\ used
in the Kast blue side''.} \vskip 6pt}
\noalign{\hbox{$^c$ These are first order values. We used second order.}\vskip
6pt}
\noalign{\hbox{$^d$ There is no useful data from 4220\AA\ down to the minimum
wavelength
of about 3800\AA.} \vskip 6pt}
\noalign{\hbox{$^e$ There are gaps between all orders -- see Table 2.}\vskip
6pt}
\noalign{\hbox{$^f$ Per binned ($2\times2$) pixels.} \vskip 6pt}
}
}$$
\vfill\eject
\pageno=25
$$\vbox{
\halign{
\tabskip=0.4em
\hskip 0.4em  \hfil#\hfil & \hfil#\hfil & \hfil#\hfil & \hfil#\hfil
            & \hfil#\hfil & \hfil#\hfil & \hfil#\hfil & \hfil#\hfil
            & \hfil#\hfil & \hfil#\hfil & \hfil#\hfil & \hfil#\hfil
            & \hfil#\hfil & \hfil#\hfil \cr
\multispan{14} \hfil TABLE 2 \hfil \cr
\multispan{14} \hfil WAVELENGTH COVERAGE OF ECHELLE SPECTRA \hfil \cr\cr
\noalign{ \hrule height .08em \vskip 2pt \hrule height .08em \vskip 6pt}
 Order  & $\lambda_{start}$ & $\lambda_{end}$ & SNR~$^a$ &\  &
 Order  & $\lambda_{start}$ & $\lambda_{end}$ & SNR~$^a$ &\  &
 Order  & $\lambda_{start}$ & $\lambda_{end}$ & SNR~$^a$  \cr
\noalign{\vskip 6pt \hrule height .08em \vskip 6pt}
 135 & 4220.0 & 4248.0 & 1  &\  & 116 & 4905.8 & 4943.6 & 16 &\  &  97 & 5866.6
& 5905.5 & 11 \cr
 134 & 4251.6 & 4279.6 & 1  &\  & 115 & 4948.5 & 4986.6 & 16 &\  &  96 & 5927.7
& 5967.0 & 11 \cr
 133 & 4283.6 & 4311.9 & 1  &\  & 114 & 4990.0 & 5030.3 & 15 &\  &  95 & 5990.1
& 6029.8 & 11 \cr
 132 & 4316.0 & 4344.3 & 2  &\  & 113 & 5036.0 & 5074.9 & 14 &\  &  94 & 6053.8
& 6094.0 & 11 \cr
 131 & 4348.8 & 4377.7 & 3  &\  & 112 & 5081.0 & 5120.1 & 15 &\  &  93 & 6118.8
& 6159.6 & 10 \cr
 130 & 4382.3 & 4411.6 & 2  &\  & 111 & 5126.8 & 5166.3 & 12 &\  &  92 & 6185.4
& 6226.5 & 11 \cr
 129 & 4416.2 & 4445.6 & 2  &\  & 110 & 5173.4 & 5213.2 & 12 &\  &  91 & 6253.3
& 6294.9 & 11 \cr
 128 & 4450.8 & 4480.3 & 3  &\  & 109 & 5220.8 & 5261.0 & 10 &\  &  90 & 6322.8
& 6364.8 & 9  \cr
 127 & 4485.8 & 4515.5 & 3  &\  & 108 & 5269.1 & 5304.1 & 11 &\  &  89 & 6394.1
& 6436.3 & 9  \cr
 126 & 4516.5 & 4551.3 & 3  &\  & 107 & 5318.4 & 5353.7 & 11 &\  &  88 & 6466.7
& 6509.4 & 10 \cr
 125 & 4552.7 & 4587.7 & 4  &\  & 106 & 5368.5 & 5404.1 & 12 &\  &  87 & 6540.3
& 6584.4 & 9  \cr
 124 & 4589.4 & 4624.7 & 6  &\  & 105 & 5419.7 & 5455.7 & 10 &\  &  86 & 6616.8
& 6660.8 & 9  \cr
 123 & 4626.7 & 4662.4 & 6  &\  & 104 & 5471.8 & 5508.1 & 12 &\  &  85 & 6694.8
& 6739.3 & 9  \cr
 122 & 4664.6 & 4700.6 & 7  &\  & 103 & 5524.9 & 5561.6 & 11 &\  &  84 & 6774.7
& 6819.0 & 8  \cr
 121 & 4703.2 & 4739.4 & 8  &\  & 102 & 5579.0 & 5616.1 & 12 &\  &  83 & 6855.9
& 6901.7 & 7  \cr
 120 & 4742.4 & 4778.9 & 9  &\  & 101 & 5634.3 & 5671.7 & 12 &\  &  82 & 6939.6
& 6985.8 & 7  \cr
 119 & 4782.2 & 4819.0 & 10 &\  & 100 & 5690.6 & 5728.4 & 12 &\  &  81 & 7025.3
& 7071.9 & 7  \cr
 118 & 4822.7 & 4859.9 & 11 &\  &  99 & 5748.1 & 5786.2 & 13 &\  &  80 & 7113.1
& 7160.4 & 9  \cr
 117 & 4863.9 & 4901.4 & 14 &\  &  98 & 5806.9 & 5845.2 & 14 &\  &  79 & 7203.0
& 7251.6 & 7  \cr
\noalign{\vskip 6pt \hrule height .08em \vskip 6pt}
\noalign{\hbox{$^a$ SNR near the center of the order, per binned pixel
of 0.07 -- 0.12~\AA\ .}\vskip 6pt}
}
}$$
\vskip 0.5cm
$$\vbox{
\halign{
\tabskip=2em
\hskip 2em # \hfil & \hfil#\hfil & \hfil#\hfil & \hfil# & \hfil#\hfil \cr
\multispan{5}\hfil {TABLE 3}\hfil\cr
\multispan{5}\hfil EMISSION LINES \hfil\cr\cr
\noalign{ \hrule height .08em \vskip 2pt\hrule height .08em \vskip 6pt}
\omit\hfil Ion \hfil & $\lambda_{\rm Lab}$ & $\lambda_{\rm obs}$
& W$_{\rm obs}$ & $z_{em}$ \cr
\noalign{\vskip 6pt \hrule height .08em \vskip 6pt}
H~I           & 1215.67  & 4946.6 & 84.6~$^a$ & 3.069 \cr
N~V           & 1240.13     & 5101.7 & 20.4  & 3.114 \cr
O~I           & 1304.46     & 5292.7 &  5.5  & 3.057 \cr
Si~IV+O~VI]   & 1397.8~$^b$  & 5652.3 & 20.0  & 3.044 \cr
C~IV          & 1549.06     & 6243.4 & 43.6  & 3.030 \cr
He~II+[O~III] & 1652.21     & 6746.5 & 11.8  & 3.083 \cr
C~III]        & 1908.73     & 7749.5 & 110.2 & 3.060 \cr
\multispan{5} \hfil $\langle z_{em} \rangle =3.051$ \hfil \cr
\noalign{\vskip 6pt \hrule height .08em \vskip 6pt}
\noalign{\hbox{$^a$ Measured from setup-C spectrum.} \vskip 5pt}
\noalign{\hbox{$^b$ See Table 8 in Tytler \& Fan (1992).} \vskip 5pt}
}
}$$
\vfill\eject
\pageno=26
$$\vbox{
\halign{
\tabskip=2em
\hskip 2em \hfil# & \hfil#\hfil & \hfil#\hfil & \hfil# & \hfil#\hfil
         & \hfil#\hfil & #\hfil & \hfil#\hfil \cr
\multispan{8}\hfil  TABLE 4 \hfil\cr
\multispan{8}\hfil ABSORPTION LINES \hfil\cr\cr
\noalign{ \hrule height .08em \vskip 2pt\hrule height .08em \vskip 6pt}
 \rm No.                       &
$\rm \lambda_{obs}$            &
$\sigma(\lambda)$              &
\omit\hfil $\rm W_{obs}$ \hfil &
$\sigma({\rm W})$              &
SNR                            &
\omit\hfil \rm ID \hfil        &
$\rm z_{abs}$                  \cr
\noalign{\vskip 6pt \hrule height 0.08em \vskip 6pt}
\multispan{8} \hfil setup-A (3241~\AA\ -- 4440~\AA) \hfil \cr
\noalign{\vskip 6pt \hrule height 0.08em \vskip 6pt}
  1 & 3541.05 &  0.30 &  2.36 &  0.41 &  6.5 & HI(920)     & 2.8449 \cr
  2 & 3549.02 &  0.24 &  2.57 &  0.36 &  7.6 & HI(923)     & 2.8445 \cr
  3 & 3554.68 &  0.26 &  1.13 &  0.26 &  8.5 &             &        \cr
  4 & 3561.10 &  0.23 &  2.72 &  0.31 &  9.5 & HI(926)     & 2.8447 \cr
  5 & 3578.18 &  0.13 &  2.79 &  0.21 & 12.9 & HI(930)     & 2.8444 \cr
    &         &       &       &       &      & HI(919)     & 2.8927 \cr
  6 & 3585.47 &  0.19 &  0.63 &  0.14 & 13.9 & HI(920)     & 2.8931 \cr
  7 & 3589.46 &  0.25 &  0.69 &  0.15 & 14.7 &             &        \cr
  8 & 3594.23 &  0.28 &  0.85 &  0.17 & 14.7 & HI(923)     & 2.8934 \cr
  9 & 3605.57 &  0.13 &  2.78 &  0.17 & 17.0 & HI(937)     & 2.8447 \cr
    &         &       &       &       &      & HI(926)     & 2.8928 \cr
 10 & 3616.06 &  0.35 &  1.06 &  0.19 & 15.7 &             &        \cr
 11 & 3623.24 &  0.17 &  1.75 &  0.17 & 16.1 & HI(930)     & 2.8928 \cr
 12 & 3628.16 &  0.12 &  1.84 &  0.15 & 16.6 &             &        \cr
 13 & 3635.43 &  0.15 &  1.03 &  0.14 & 16.0 &             &        \cr
 14 & 3639.63 &  0.10 &  2.14 &  0.15 & 16.6 & CI(1329)    & 1.7390 \cr
 15 & 3652.16 &  0.10 &  4.65 &  0.24 & 12.3 & HI(949)     & 2.8454 \cr
    &         &       &       &       &      & HI(937)     & 2.8944 \cr
 16 & 3658.44 &  0.19 &  2.52 &  0.21 & 15.2 &             &        \cr
 17 & 3673.64 &  0.15 &  2.41 &  0.16 & 19.5 &             &        \cr
 18 & 3681.45 &  0.18 &  1.30 &  0.13 & 20.2 &             &        \cr
 19 & 3689.50 &  0.13 &  2.61 &  0.15 & 21.2 &             &        \cr
 20 & 3697.49 &  0.08 &  1.99 &  0.12 & 21.2 & HI(949)     & 2.8931 \cr
 21 & 3701.25 &  0.13 &  2.21 &  0.14 & 20.8 &             &        \cr
 22 & 3709.40 &  0.21 &  0.81 &  0.12 & 20.6 &             &        \cr
 23 & 3727.01 &  0.19 &  2.34 &  0.15 & 22.7 &             &        \cr
 24 & 3738.26 &  0.07 &  5.02 &  0.13 & 25.3 & HI(972)     & 2.8438 \cr
    &         &       &       &       &      & HI(923)     & 3.0495 \cr
 25 & 3751.31 &  0.08 &  2.35 &  0.10 & 26.5 & HI(926)     & 3.0501 \cr
    &         &       &       &       &      & NiII(1370)  & 1.7379 \cr
 26 & 3755.88 &  0.05 &  2.63 &  0.09 & 27.7 & CIII(977)   & 2.8442 \cr
 27 & 3762.85 &  0.09 &  2.74 &  0.11 & 27.8 &             &        \cr
 28 & 3769.66 &  0.06 &  1.55 &  0.08 & 28.9 & HI(930)     & 3.0501 \cr
 29 & 3778.34 &  0.09 &  1.77 &  0.09 & 28.8 &             &        \cr
 30 & 3785.92 &  0.07 &  2.66 &  0.10 & 30.5 & HI(972)     & 2.8926 \cr
 31 & 3798.23 &  0.06 &  2.51 &  0.09 & 31.8 & HI(937)     & 3.0501 \cr
\noalign{\vskip 6pt \hrule height 0.08em }}
}$$
\vfill\eject
$$\vbox{\halign{
\tabskip=2em
\hskip 2em  \hfil# & \hfil#\hfil & \hfil # \hfil & \hfil#
& \hfil # \hfil & \hfil # \hfil & #\hfil & \hfil#\hfil \cr
\multispan{8}\hfil  TABLE 4--{\it continued}  \hfil\cr\cr
\noalign{ \hrule height .08em \vskip 2pt\hrule height .08em \vskip 6pt}
 \rm No.                       &
$\rm \lambda_{obs}$            &
$\sigma(\lambda)$              &
\omit\hfil $\rm W_{obs}$ \hfil &
$\sigma({\rm W})$              &
SNR                            &
\omit\hfil \rm ID \hfil        &
$\rm z_{abs}$                  \cr
\noalign{\vskip 6pt \hrule height 0.08em \vskip 6pt}
 32 & 3803.05 &  0.08 &  1.53 &  0.08 & 30.2 & CIII(977)   & 2.8932 \cr
 33 & 3808.72 &  0.26 &  0.48 &  0.08 & 29.1 &             &        \cr
 34 & 3817.34 &  0.10 &  4.98 &  0.13 & 31.4 & SiIV(1393)  & 1.7389 \cr
 35 & 3834.02 &  0.12 &  1.03 &  0.08 & 29.4 &             &        \cr
 36 & 3840.36 &  0.21 &  0.75 &  0.09 & 30.0 & SiIV(1402)  & 1.7377 \cr
 37 & 3845.93 &  0.07 &  2.18 &  0.08 & 31.7 & HI(949)     & 3.0494 \cr
 38 & 3856.82 &  0.08 &  1.82 &  0.08 & 32.8 &             &        \cr
 39 & 3863.01 &  0.04 &  3.57 &  0.07 & 35.8 &             &        \cr
 40 & 3874.59 &  0.07 &  3.11 &  0.09 & 34.3 &             &        \cr
 41 & 3897.53 &  0.07 &  1.53 &  0.07 & 34.5 &             &        \cr
 42 & 3905.93 &  0.07 &  2.13 &  0.08 & 32.8 &             &        \cr
 43 & 3912.42 &  0.09 &  0.99 &  0.07 & 29.7 &             &        \cr
 44 & 3925.22 &  0.09 &  3.66 &  0.10 & 36.4 &             &        \cr
 45 & 3933.80 &  0.04 &  3.02 &  0.07 & 39.4 &             &        \cr
 46 & 3938.03 &  0.02 &  2.77 &  0.05 & 42.9 & HI(972)     & 3.0492 \cr
 47 & 3943.33 &  0.03 &  4.76 &  0.07 & 40.2 & HI(1025)    & 2.8444 \cr
 48 & 3956.43 &  0.06 &  2.13 &  0.07 & 37.9 & CIII(977)   & 3.0495 \cr
 49 & 3967.52 &  0.08 &  3.22 &  0.09 & 40.1 & OVI(1031)   & 2.8448 \cr
 50 & 3984.19 &  0.08 &  1.23 &  0.07 & 37.8 & CII(1036)   & 2.8445 \cr
 51 & 3989.00 &  0.03 &  2.09 &  0.05 & 41.5 & OVI(1037)   & 2.8444 \cr
 52 & 3992.78 &  0.03 &  2.87 &  0.06 & 42.5 & HI(1025)    & 2.8927 \cr
 53 & 4002.90 &  0.15 &  0.30 &  0.05 & 36.4 & OI(988)     & 3.0491 \cr
 54 & 4008.85 &  0.16 &  1.02 &  0.08 & 37.0 & SiII(989)   & 3.0499 \cr
    &         &       &       &       &      & NIII(989)   & 3.0502 \cr
 55 & 4017.29 &  0.12 &  1.30 &  0.08 & 36.8 & OVI(1031)   & 2.8930 \cr
 56 & 4030.70 &  0.08 &  2.71 &  0.09 & 38.0 &             &        \cr
 57 & 4040.90 &  0.20 &  0.48 &  0.07 & 36.5 &             &        \cr
 58 & 4045.84 &  0.25 &  0.42 &  0.09 & 26.4 &             &        \cr
 59 & 4050.24 &  0.17 &  1.00 &  0.10 & 29.2 &             &        \cr
 60 & 4072.49 &  0.17 &  0.94 &  0.08 & 38.1 &             &        \cr
 61 & 4090.06 &  0.09 &  3.09 &  0.09 & 40.3 &             &        \cr
 62 & 4104.55 &  0.44 &  0.42 &  0.08 & 39.2 &             &        \cr
 63 & 4111.17 &  0.22 &  0.55 &  0.07 & 39.5 &             &        \cr
 64 & 4118.08 &  0.13 &  0.86 &  0.07 & 40.2 &             &        \cr
 65 & 4127.98 &  0.07 &  3.22 &  0.08 & 41.9 &             &        \cr
 66 & 4155.34 &  0.07 &  4.40 &  0.09 & 42.5 & HI(1025)    & 3.0511 \cr
 67 & 4168.46 &  0.27 &  0.56 &  0.08 & 38.2 & NII(1083)   & 2.8455 \cr
 68 & 4179.05 &  0.07 &  3.36 &  0.08 & 42.6 & OVI(1031)   & 3.0498 \cr
    &         &       &       &       &      & SiII(1526)  & 1.7373 \cr
\noalign{\vskip 6pt \hrule height 0.08em }}
}$$
\vfill\eject
$$\vbox{\halign{
\tabskip=2em
\hskip 2em  \hfil# & \hfil#\hfil & \hfil # \hfil & \hfil#
& \hfil # \hfil & \hfil # \hfil & #\hfil & \hfil#\hfil \cr
\multispan{8}\hfil  TABLE 4--{\it continued}  \hfil\cr\cr
\noalign{ \hrule height .08em \vskip 2pt\hrule height .08em \vskip 6pt}
 \rm No.                       &
$\rm \lambda_{obs}$            &
$\sigma(\lambda)$              &
\omit\hfil $\rm W_{obs}$ \hfil &
$\sigma({\rm W})$              &
SNR                            &
\omit\hfil \rm ID \hfil        &
$\rm z_{abs}$                  \cr
\noalign{\vskip 6pt \hrule height 0.08em \vskip 6pt}
 69 & 4195.08 &  0.10 &  1.78 &  0.07 & 42.2 &             &        \cr
 70 & 4201.53 &  0.08 &  2.35 &  0.08 & 43.3 & OVI(1037)   & 3.0492 \cr
 71 & 4213.55 &  0.08 &  3.32 &  0.08 & 42.9 &             &        \cr
 72 & 4234.65 &  0.26 &  1.40 &  0.10 & 40.9 &             &        \cr
 73 & 4246.77 &  0.25 &  0.73 &  0.08 & 39.6 & CIV(1550)   & 1.7385 \cr
 74 & 4254.32 &  0.14 &  2.28 &  0.09 & 41.6 &             &        \cr
 75 & 4275.10 &  0.14 &  2.56 &  0.09 & 41.9 &             &        \cr
 76 & 4294.83 &  0.14 &  1.69 &  0.08 & 42.0 &             &        \cr
 77 & 4300.33 &  0.09 &  5.32 &  0.10 & 44.1 &             &        \cr
 78 & 4333.87 &  0.13 &  2.80 &  0.10 & 41.0 &             &        \cr
 79 & 4346.99 &  0.15 &  1.30 &  0.08 & 36.8 &             &        \cr
 80 & 4357.22 &  0.15 &  2.32 &  0.13 & 28.0 &             &        \cr
 81 & 4396.12 &  0.08 &  2.00 &  0.08 & 37.7 &             &        \cr
 82 & 4404.79 &  0.15 &  2.55 &  0.11 & 35.8 & FeII(1608)  & 1.7384 \cr
\noalign{\vskip 6pt \hrule height 0.08em \vskip 6pt}
\multispan{8} \hfil setup-B (4045~\AA\ -- 4963~\AA) \hfil \cr
\noalign{\vskip 6pt \hrule height 0.08em \vskip 6pt}
  1 & 4059.13 &  0.20 &  0.27 &  0.05 & 37.7 &             &        \cr
  2 & 4065.45 &  0.17 &  0.37 &  0.04 & 44.4 &             &        \cr
  3 & 4073.39 &  0.12 &  1.11 &  0.05 & 50.7 &             &        \cr
  4 & 4090.52 &  0.08 &  2.57 &  0.05 & 57.1 &             &        \cr
  5 & 4105.70 &  0.36 &  0.22 &  0.04 & 57.5 &             &        \cr
  6 & 4111.74 &  0.10 &  0.35 &  0.03 & 58.7 &             &        \cr
  7 & 4118.46 &  0.08 &  0.78 &  0.03 & 59.5 &             &        \cr
  8 & 4127.85 &  0.07 &  2.82 &  0.05 & 61.1 &             &        \cr
  9 & 4153.79 &  0.04 &  2.41 &  0.04 & 63.4 & HI(1025)    & 3.0496 \cr
 10 & 4158.06 &  0.05 &  1.79 &  0.04 & 62.0 &             &        \cr
 11 & 4167.32 &  0.15 &  0.34 &  0.03 & 54.6 & NII(1083)   & 2.8444 \cr
 12 & 4178.59 &  0.06 &  2.97 &  0.05 & 59.5 & OVI(1031)   & 3.0493 \cr
 13 & 4194.82 &  0.05 &  1.18 &  0.04 & 57.2 &             &        \cr
 14 & 4201.99 &  0.06 &  1.67 &  0.04 & 58.3 & OVI(1037)   & 3.0497 \cr
 15 & 4209.00 &  0.07 &  0.75 &  0.03 & 54.8 &             &        \cr
 16 & 4213.65 &  0.06 &  2.62 &  0.05 & 56.5 &             &        \cr
 17 & 4234.56 &  0.16 &  0.80 &  0.05 & 53.0 &             &        \cr
 18 & 4239.77 &  0.16 &  0.32 &  0.04 & 52.2 & CIV(1548)   & 1.7385 \cr
 19 & 4247.31 &  0.20 &  0.46 &  0.05 & 49.4 & CIV(1550)   & 1.7388 \cr
 20 & 4254.55 &  0.06 &  1.70 &  0.05 & 53.4 &             &        \cr
 21 & 4261.62 &  0.11 &  0.48 &  0.04 & 51.9 &             &        \cr
 22 & 4275.37 &  0.07 &  1.28 &  0.04 & 52.5 &             &        \cr
\noalign{\vskip 6pt \hrule height 0.08em }}
}$$
\vfill\eject
$$\vbox{\halign{
\tabskip=2em
\hskip 2em \hfil# & \hfil#\hfil & \hfil # \hfil & \hfil#
& \hfil # \hfil & \hfil # \hfil & #\hfil & \hfil#\hfil \cr
\multispan{8}\hfil  TABLE 4--{\it continued}  \hfil\cr\cr
\noalign{ \hrule height .08em \vskip 2pt\hrule height .08em \vskip 6pt}
 \rm No.                       &
$\rm \lambda_{obs}$            &
$\sigma(\lambda)$              &
\omit\hfil $\rm W_{obs}$ \hfil &
$\sigma({\rm W})$              &
SNR                            &
\omit\hfil \rm ID \hfil        &
$\rm z_{abs}$                  \cr
\noalign{\vskip 6pt \hrule height 0.08em \vskip 6pt}
 23 & 4292.55 &  0.09 &  0.82 &  0.04 & 50.9 &             &        \cr
 24 & 4299.97 &  0.05 &  4.13 &  0.05 & 54.7 &             &        \cr
 25 & 4306.79 &  0.04 &  0.98 &  0.03 & 51.5 &             &        \cr
 26 & 4309.77 &  0.10 &  0.73 &  0.04 & 48.8 &             &        \cr
 27 & 4321.94 &  0.18 &  0.24 &  0.04 & 45.7 &             &        \cr
 28 & 4329.75 &  0.09 &  0.78 &  0.04 & 49.2 &             &        \cr
 29 & 4334.78 &  0.05 &  1.62 &  0.04 & 51.6 &             &        \cr
 30 & 4347.18 &  0.12 &  1.60 &  0.06 & 45.8 &             &        \cr
 31 & 4355.53 &  0.05 &  1.91 &  0.05 & 42.9 &             &        \cr
 32 & 4381.59 &  0.13 &  0.46 &  0.04 & 45.3 &             &        \cr
 33 & 4390.95 &  0.09 &  0.84 &  0.04 & 46.4 & NII(1083)   & 3.0507 \cr
 34 & 4396.12 &  0.05 &  2.01 &  0.05 & 47.7 &             &        \cr
 35 & 4404.03 &  0.09 &  1.61 &  0.06 & 44.3 & FeII(1608)  & 1.7381 \cr
 36 & 4430.05 &  0.03 &  3.43 &  0.05 & 45.9 &             &        \cr
 37 & 4434.73 &  0.03 &  2.41 &  0.05 & 44.0 &             &        \cr
 38 & 4441.98 &  0.08 &  3.49 &  0.07 & 42.0 &             &        \cr
 39 & 4449.87 &  0.10 &  0.64 &  0.05 & 41.2 &             &        \cr
 40 & 4460.46 &  0.09 &  0.72 &  0.04 & 42.5 &             &        \cr
 41 & 4463.73 &  0.09 &  0.51 &  0.04 & 42.3 &             &        \cr
 42 & 4468.90 &  0.14 &  0.31 &  0.04 & 41.0 &             &        \cr
 43 & 4476.81 &  0.15 &  0.60 &  0.04 & 41.0 &             &        \cr
 44 & 4486.93 &  0.06 &  2.11 &  0.06 & 43.2 &             &        \cr
 45 & 4499.67 &  0.11 &  1.05 &  0.05 & 41.1 &             &        \cr
 46 & 4505.76 &  0.07 &  0.68 &  0.04 & 41.4 &             &        \cr
 47 & 4509.92 &  0.18 &  0.62 &  0.06 & 40.8 &             &        \cr
 48 & 4521.24 &  0.15 &  0.44 &  0.05 & 40.6 &             &        \cr
 49 & 4531.26 &  0.10 &  0.65 &  0.05 & 40.8 &             &        \cr
 50 & 4536.75 &  0.05 &  0.36 &  0.03 & 41.2 & CI(1656)    & 1.7380 \cr
 51 & 4543.98 &  0.14 &  2.84 &  0.08 & 40.3 &             &        \cr
 52 & 4573.57 &  0.10 &  3.93 &  0.08 & 41.9 & SiII(1190)  & 2.8420 \cr
    &         &       &       &       &      & AlII(1670)  & 1.7376 \cr
 53 & 4587.31 &  0.07 &  1.33 &  0.05 & 41.6 & SiII(1193)  & 2.8442 \cr
 54 & 4592.02 &  0.04 &  2.90 &  0.06 & 44.0 &             &        \cr
 55 & 4601.21 &  0.29 &  0.24 &  0.04 & 39.8 &             &        \cr
 56 & 4608.49 &  0.11 &  0.29 &  0.04 & 40.2 &             &        \cr
 57 & 4619.64 &  0.07 &  0.43 &  0.04 & 41.1 &             &        \cr
 58 & 4625.69 &  0.18 &  0.54 &  0.04 & 39.2 &             &        \cr
 59 & 4632.57 &  0.10 &  0.48 &  0.04 & 39.1 &             &        \cr
\noalign{\vskip 6pt \hrule height 0.08em }}
}$$
\vfill\eject
$$\vbox{\halign{
\tabskip=2em
\hskip 2em \hfil# & \hfil#\hfil & \hfil # \hfil & \hfil#
& \hfil # \hfil & \hfil # \hfil & #\hfil & \hfil#\hfil \cr
\multispan{8}\hfil  TABLE 4--{\it continued}  \hfil\cr\cr
\noalign{ \hrule height .08em \vskip 2pt\hrule height .08em \vskip 6pt}
 \rm No.                       &
$\rm \lambda_{obs}$            &
$\sigma(\lambda)$              &
\omit\hfil $\rm W_{obs}$ \hfil &
$\sigma({\rm W})$              &
SNR                            &
\omit\hfil \rm ID \hfil        &
$\rm z_{abs}$                  \cr
\noalign{\vskip 6pt \hrule height 0.08em \vskip 6pt}
 60 & 4637.89 &  0.03 &  2.52 &  0.05 & 43.0 & SiIII(1206) & 2.8441 \cr
 61 & 4641.27 &  0.03 &  1.47 &  0.04 & 41.4 &             &        \cr
 62 & 4646.87 &  0.12 &  0.61 &  0.05 & 37.6 &             &        \cr
 63 & 4654.25 &  0.07 &  1.80 &  0.06 & 38.3 &             &        \cr
 64 & 4673.85 &  0.11 & 21.67 &  0.16 & 40.7 & HI(1215)    & 2.8447 \cr
 65 & 4692.04 &  0.05 &  1.52 &  0.05 & 43.9 &             &        \cr
 66 & 4696.56 &  0.08 &  0.55 &  0.04 & 43.5 & SiIII(1206) & 2.8927 \cr
 67 & 4706.89 &  0.09 &  0.83 &  0.04 & 46.9 &             &        \cr
 68 & 4714.05 &  0.30 &  0.35 &  0.04 & 47.7 &             &        \cr
 69 & 4721.58 &  0.04 &  2.07 &  0.04 & 54.8 &             &        \cr
 70 & 4731.84 &  0.04 &  2.87 &  0.05 & 56.5 & HI(1215)    & 2.8924 \cr
 71 & 4743.30 &  0.21 &  0.42 &  0.04 & 50.5 &             &        \cr
 72 & 4749.12 &  0.10 &  0.36 &  0.03 & 53.8 &             &        \cr
 73 & 4752.57 &  0.07 &  0.97 &  0.04 & 53.8 &             &        \cr
 74 & 4761.17 &  0.03 &  3.65 &  0.04 & 63.4 &             &        \cr
 75 & 4765.97 &  0.10 &  0.64 &  0.04 & 58.9 &             &        \cr
 76 & 4774.07 &  0.05 &  1.19 &  0.03 & 62.0 &             &        \cr
 77 & 4779.54 &  0.07 &  0.65 &  0.03 & 60.2 &             &        \cr
 78 & 4793.87 &  0.05 &  1.19 &  0.03 & 66.2 &             &        \cr
 79 & 4799.32 &  0.17 &  0.33 &  0.03 & 64.4 &             &        \cr
 80 & 4810.63 &  0.16 &  0.26 &  0.03 & 65.6 &             &        \cr
 81 & 4818.61 &  0.07 &  0.44 &  0.03 & 67.1 &             &        \cr
 82 & 4823.92 &  0.08 &  0.37 &  0.03 & 66.6 &             &        \cr
 83 & 4831.30 &  0.13 &  0.32 &  0.03 & 66.4 &             &        \cr
 84 & 4835.64 &  0.04 &  0.64 &  0.02 & 67.9 &             &        \cr
 85 & 4843.69 &  0.03 &  3.53 &  0.04 & 76.5 & SiII(1260)  & 2.8429  \cr
 86 & 4853.54 &  0.03 &  1.83 &  0.03 & 75.9 &             &        \cr
 87 & 4858.16 &  0.05 &  0.52 &  0.02 & 73.2 &             &        \cr
 88 & 4863.53 &  0.10 &  0.63 &  0.03 & 71.2 &             &        \cr
 89 & 4872.32 &  0.03 &  2.39 &  0.03 & 80.1 &             &        \cr
 90 & 4880.93 &  0.04 &  1.10 &  0.03 & 78.3 &             &        \cr
 91 & 4885.70 &  0.02 &  1.82 &  0.02 & 88.2 & SiIII(1206) & 3.0495 \cr
 92 & 4889.39 &  0.03 &  1.70 &  0.02 & 85.3 &             &        \cr
 93 & 4896.90 &  0.02 &  3.59 &  0.03 & 92.2 &             &        \cr
 94 & 4904.38 &  0.04 &  1.68 &  0.03 & 84.9 &             &        \cr
 95 & 4909.49 &  0.02 &  1.51 &  0.02 & 89.3 &             &        \cr
 96 & 4918.81 &  0.04 &  0.80 &  0.02 & 82.1 &             &        \cr
 97 & 4922.40 &  0.03 &  3.75 &  0.03 & 98.5 & HI(1215)    & 3.0491 \cr
\noalign{\vskip 6pt \hrule height .08em \vskip 6pt}
}
}$$
\vfill\eject
$$\vbox{\halign{
\tabskip=2em
\hskip 2em \hfil# & \hfil#\hfil & \hfil # \hfil & \hfil#
& \hfil # \hfil & \hfil # \hfil & #\hfil & \hfil#\hfil \cr
\multispan{8}\hfil  TABLE 4--{\it continued}  \hfil\cr\cr
\noalign{ \hrule height .08em \vskip 2pt\hrule height .08em \vskip 6pt}
 \rm No.                       &
$\rm \lambda_{obs}$            &
$\sigma(\lambda)$              &
\omit\hfil $\rm W_{obs}$ \hfil &
$\sigma({\rm W})$              &
SNR                            &
\omit\hfil \rm ID \hfil        &
$\rm z_{abs}$                  \cr
\noalign{\vskip 6pt \hrule height 0.08em \vskip 6pt}
\multispan{8} \hfil setup-C (4817~\AA\ -- 6718~\AA) \hfil \cr
\noalign{\vskip 6pt \hrule height 0.08em \vskip 6pt}
  1 & 5006.45 &  0.26 &  0.31 &  0.03 & 82.0 & OI(1302)    & 2.8447 \cr
  2 & 5017.29 &  0.26 &  0.49 &  0.04 & 83.7 & NV(1238)    & 3.0501 \cr
  3 & 5033.63 &  0.50 &  0.16 &  0.03 & 81.9 & NV(1242)    & 3.0502 \cr
  4 & 5078.85 &  0.32 &  0.15 &  0.03 & 80.8 & AlIII(1854) & 1.7383 \cr
  5 & 5104.76 &  0.24 &  0.47 &  0.04 & 81.8 & SiII(1260)  & 3.0500 \cr
  6 & 5131.92 &  0.14 &  0.58 &  0.03 & 83.0 & CII(1334)   & 2.8455 \cr
  7 & 5357.90 &  0.05 &  1.42 &  0.03 & 75.5 & SiIV(1393)  & 2.8442 \cr
  8 & 5392.48 &  0.10 &  0.98 &  0.04 & 77.7 & SiIV(1402)  & 2.8442 \cr
  9 & 5404.82 &  0.20 &  0.50 &  0.04 & 78.4 & CII(1334)   & 3.0500 \cr
 10 & 5630.94 &  0.54 &  0.14 &  0.03 & 75.0 & CrII(2056)  & 1.7384 \cr
 11 & 5644.59 &  0.05 &  2.70 &  0.04 & 80.4 & SiIV(1393)  & 3.0499 \cr
 12 & 5651.62 &  0.27 &  0.13 &  0.03 & 72.5 & CIV(1550)   & 2.6444 \cr
 13 & 5657.44 &  0.19 &  0.17 &  0.03 & 70.7 & CrII(2066)  & 1.7381 \cr
 14 & 5680.98 &  0.07 &  1.96 &  0.06 & 51.9 & SiIV(1402)  & 3.0498 \cr
 15 & 5869.54 &  0.33 &  0.37 &  0.06 & 43.9 & SiII(1526)  & 2.8446 \cr
 16 & 5925.46 &  0.19 &  0.50 &  0.05 & 45.2 & MgII(2796)  & 1.1190 \cr
 17 & 5940.83 &  0.22 &  0.45 &  0.06 & 43.3 & MgII(2803)  & 1.1191 \cr
 18 & 5951.30 &  0.13 &  1.85 &  0.08 & 43.8 & CIV(1548)   & 2.8440 \cr
 19 & 5961.05 &  0.08 &  1.13 &  0.05 & 45.3 & CIV(1550)   & 2.8439 \cr
 20 & 6026.62 &  0.24 &  0.35 &  0.05 & 50.9 & CIV(1548)   & 2.8927 \cr
 21 & 6037.24 &  0.38 &  0.24 &  0.05 & 51.9 & CIV(1550)   & 2.8930 \cr
\noalign{\vskip 6pt \hrule height 0.08em \vskip 6pt}
\multispan{8} \hfil setup-D (5990~\AA\ -- 8798~\AA) \hfil \cr
\noalign{\vskip 6pt \hrule height 0.08em \vskip 6pt}
  1 & 6026.96 &  0.22 &  0.29 &  0.06 & 84.6 & CIV(1548)   & 2.8929 \cr
  2 & 6036.18 &  0.29 &  0.23 &  0.05 & 85.8 & CIV(1550)   & 2.8924 \cr
  3 & 6270.26 &  0.03 &  4.35 &  0.06 &113.4 & CIV(1548)   & 3.0500 \cr
  4 & 6280.60 &  0.02 &  3.88 &  0.05 &119.6 & CIV(1550)   & 3.0500 \cr
  5 & 6420.75 &  0.16 &  1.28 &  0.07 &105.8 & FeII(2344)  & 1.7390 \cr
  6 & 6502.14 &  0.16 &  0.22 &  0.04 & 94.9 & FeII(2374)  & 1.7384 \cr
  7 & 6524.55 &  0.14 &  0.85 &  0.06 &104.0 & FeII(2382)  & 1.7382 \cr
  8 & 7083.24 &  0.13 &  0.56 &  0.05 &104.0 & FeII(2586)  & 1.7384 \cr
  9 & 7119.83 &  0.08 &  0.65 &  0.04 &108.7 & FeII(2600)  & 1.7382 \cr
 10 & 7290.39 &  0.13 &  1.09 &  0.07 & 87.3 &             &        \cr
 11 & 7303.33 &  0.20 &  0.83 &  0.07 & 93.0 &             &        \cr
 12 & 7341.28 &  0.22 &  1.04 &  0.09 & 83.5 &             &        \cr
 13 & 7655.59 &  0.09 &  0.97 &  0.06 & 85.4 & MgII(2796)  & 1.7377 \cr
 14 & 7676.63 &  0.09 &  1.39 &  0.06 & 97.9 & MgII(2803)  & 1.7382 \cr
\noalign{\vskip 6pt \hrule height .08em \vskip 6pt}
}
}$$
\vfill\eject
\pageno=32
$$\vbox{
\halign{
\tabskip=2em
\hskip 2em \hfil # & # \hfil & # \hfil & \hfil#\hfil & \hfil#\hfil
& \hfil#\hfil &\hfil #  \hfil \cr
\multispan{7}\hfil TABLE 5 \hfil \cr
\multispan{7}\hfil RESULTS FROM VOIGT PROFILE FITS\hfil\cr\cr
\noalign{ \hrule height .08em \vskip 2pt \hrule height .08em \vskip 6pt}
No. & \omit\hfil \rm Cross-ID \hfil  &  Line ID. & $\rm \lambda_{vac}^{a}$
& $\rm z_{abs}$ &  log $N$ (cm$^{-2}$) &  $b$ (km~s$^{-1}$) \cr
\noalign{\vskip 6pt \hrule height .08em \vskip 6pt}
  1 & B--20 & \Lya & 4255.45 & 2.5005 & 17.16$\pm$3.00 & 10 \cr
  2 &       & \Lya & 4267.85 & 2.5107 & 13.58$\pm$0.66 & 32 \cr
  3 &       & \Lya & 4269.55 & 2.5121 & 13.47$\pm$0.60 & 25 \cr
  4 & B--22 & \Lya & 4275.15 & 2.5167 & 13.58$\pm$0.65 & 14 \cr
  5 &       & \Lya & 4290.71 & 2.5295 & 13.91$\pm$1.04 & 15 \cr
  6 & B--23 & \Lya & 4292.77 & 2.5312 & 16.33$\pm$1.60 & 18 \cr
  7 &       & \Lya & 4294.60 & 2.5327 & 13.30$\pm$0.92 & 11 \cr
  8 &       & \Lya & 4296.54 & 2.5343 & 13.66$\pm$0.88 & 21 \cr
  9 & B--24 & \Lya & 4298.37 & 2.5358 & 13.79$\pm$0.67 & 11 \cr
 10 & B--24 & \Lya & 4299.09 & 2.5364 & 13.78$\pm$0.51 & 17 \cr
 11 & B--24 & \Lya & 4300.19 & 2.5373 & 13.74$\pm$0.56 & 14 \cr
 12 & B--24 & \Lya & 4300.92 & 2.5379 & 13.73$\pm$0.49 & 19 \cr
 13 &       & \Lya & 4303.72 & 2.5402 & 13.68$\pm$0.59 & 26 \cr
 14 & B--25 & \Lya & 4307.00 & 2.5429 & 13.71$\pm$0.77 & 17 \cr
 15 & B--28 & \Lya & 4329.85 & 2.5617 & 13.85$\pm$0.51 & 33 \cr
 16 & B--29 & \Lya & 4334.11 & 2.5652 & 16.09$\pm$1.88 & 13 \cr
 17 &       & \Lya & 4335.20 & 2.5661 & 14.06$\pm$0.76 & 37 \cr
 18 &       & \Lya & 4374.47 & 2.5984 & 13.92$\pm$1.50 & 13 \cr
 19 & B--36 & \Lya & 4396.11 & 2.6162 & 14.10$\pm$0.56 & 44 \cr
 20 & B--36 & \Lya & 4397.08 & 2.6170 & 13.76$\pm$1.12 & 19 \cr
    &       & Si~III(1206) & 4397.09 & 2.6445 & 13.16$\pm$0.89 & 19 \cr
 21 & B--37 & \Lya & 4403.76 & 2.6225 & 14.02$\pm$0.53 & 44 \cr
    && Fe~II(1608) & 4403.78 & 1.7379 & 14.73$\pm$0.51 & 45 \cr
 22 &       & \Lya & 4421.03 & 2.6367 & 13.97$\pm$0.78 & 51 \cr
 23 &       & \Lya & 4422.24 & 2.6377 & 16.92$\pm$3.77 & 11 \cr
 24 &       & \Lya & 4424.67 & 2.6397 & 13.56$\pm$0.39 & 18 \cr
 25 &       & \Lya & 4425.28 & 2.6402 & 13.52$\pm$0.47 & 12 \cr
 26 & B--36 & \Lya & 4428.56 & 2.6429 & 14.76$\pm$0.73 & 37 \cr
 27 & B--36 & \Lya & 4430.14 & 2.6442 & 17.37$\pm$0.56 &  7 \cr
 28 & B--36 & \Lya & 4431.24 & 2.6451 & 13.94$\pm$0.43 & 39 \cr
 29 & B--37 & \Lya & 4434.28 & 2.6476 & 14.30$\pm$1.06 & 36 \cr
 30 & B--37 & \Lya & 4435.74 & 2.6488 & 14.75$\pm$0.66 & 18 \cr
 31 &       & \Lya & 4439.14 & 2.6516 & 13.51$\pm$0.40 & 14 \cr
 32 & B--38 & \Lya & 4441.45 & 2.6535 & 14.79$\pm$1.04 & 14 \cr
 33 & B--38 & \Lya & 4442.67 & 2.6545 & 14.48$\pm$0.77 & 14 \cr
 34 &       & \Lya & 4455.43 & 2.6650 & 13.76$\pm$0.60 & 46 \cr
 35 & B--40 & \Lya & 4460.05 & 2.6688 & 14.01$\pm$0.77 & 22 \cr
\noalign{\vskip 6pt \hrule height .08em \vskip 6pt}
}
}$$
\vfill\eject
$$\vbox{
\halign{
\tabskip=2em
\hskip 2em \hfil # & # \hfil & # \hfil & \hfil#\hfil & \hfil#\hfil
& \hfil#\hfil &\hfil # \hfil \cr
\multispan{7}\hfil TABLE 5 --- {\it continued} \hfil \cr\cr
\noalign{ \hrule height .08em \vskip 2pt \hrule height .08em \vskip 6pt}
No. & \omit\hfil \rm Cross-ID \hfil  &  Line ID. & $\rm \lambda_{vac}^{a}$
& $\rm z_{abs}$ &  log $N$ (cm$^{-2}$) &  $b$ (km~s$^{-1}$)  \cr
\noalign{\vskip 6pt \hrule height .08em \vskip 6pt}
 36 & B--41 & \Lya & 4464.18 & 2.6722 & 13.64$\pm$0.45 & 18 \cr
 37 & B--43 & \Lya & 4476.10 & 2.6820 & 13.41$\pm$0.69 & 27 \cr
 38 & B--43 & \Lya & 4477.31 & 2.6830 & 13.74$\pm$0.58 & 28 \cr
 39 & B--45 & \Lya & 4499.68 & 2.7014 & 13.51$\pm$0.55 & 22 \cr
 40 & B--45 & \Lya & 4500.53 & 2.7021 & 14.05$\pm$1.05 & 10 \cr
 41 & B--46 & \Lya & 4505.64 & 2.7063 & 13.96$\pm$0.38 & 51 \cr
 42 & B--46 & \Lya & 4506.00 & 2.7066 & 14.44$\pm$2.90 & 11 \cr
 43 & B--47 & \Lya & 4509.65 & 2.7096 & 13.49$\pm$0.55 & 27 \cr
 44 & B--47 & \Lya & 4510.62 & 2.7104 & 13.61$\pm$0.46 & 25 \cr
 45 &       & \Lya & 4513.18 & 2.7125 & 13.51$\pm$0.63 & 25 \cr
 46 &       & \Lya & 4528,37 & 2.7250 & 13.61$\pm$1.13 & 17 \cr
 47 & B--49 & \Lya & 4531.05 & 2.7272 & 13.64$\pm$0.51 & 32 \cr
 48 & B--49 & \Lya & 4531.90 & 2.7279 & 13.65$\pm$0.67 & 25 \cr
 49 & B--50 & \Lya & 4537.25 & 2.7323 & 13.68$\pm$0.50 & 58 \cr
    & & C~I(1656)  & 4537.33 & 1.7384 & 14.26$\pm$0.46 & 58 \cr
 50 &       & \Lya & 4541.01 & 2.7354 & 13.63$\pm$0.47 & 53 \cr
 51 & B--51 & \Lya & 4544.05 & 2.7379 & 14.18$\pm$0.30 & 64 \cr
 52 & B--52 & \Lya & 4570.68 & 2.7598 & 13.78$\pm$0.53 & 30 \cr
 53 & B--52 & \Lya & 4572.50 & 2.7613 & 14.05$\pm$0.60 & 29 \cr
 54 & B--52 & Al~II(1670) & 4574.78 & 1.7381 & 13.35$\pm$0.59 & 33 \cr
 55 & B--52 & Si~II(1190) & 4576.32 & 2.8443 & 13.60$\pm$0.60 & 10 \cr
 56 & B--52 & \Lya & 4577.97 & 2.7658 & 13.55$\pm$0.56 & 34 \cr
 57 & B--53 &Si~II(1193) & 4587.13 & 2.8441 & 13.28$\pm$0.45 & 12 \cr
 58 & B--54 & \Lya & 4592.07 & 2.7774 & 14.21$\pm$0.94 & 39 \cr
 59 & B--54 & \Lya & 4593.41 & 2.7785 & 13.80$\pm$1.00 & 27 \cr
    &       & Si~II(1260) & 3593.36 & 2.6443 & 13.43$\pm$0.92 & 27 \cr
 60 & B--55 & \Lya & 4601.80 & 2.7854 & 13.54$\pm$0.61 & 27 \cr
 61 & B--55 & \Lya & 4602.89 & 2.7863 & 13.03$\pm$1.01 & 17 \cr
 62 & B--56 & \Lya & 4608.60 & 2.7910 & 13.46$\pm$0.45 & 33 \cr
 63 & B--57 & \Lya & 4619.55 & 2.8000 & 13.37$\pm$0.55 & 17 \cr
 64 & B--57 & \Lya & 4620.28 & 2.8006 & 13.65$\pm$0.34 & 26 \cr
 65 &       & \Lya & 4622.22 & 2.8022 & 12.93$\pm$0.47 & 15 \cr
 66 & B--59 & \Lya & 4633.28 & 2.8113 & 13.59$\pm$0.36 & 26 \cr
 67 & B--60 & Si~III(1206) & 4636.94 & 2.8433 & 13.34$\pm$0.58 & 42 \cr
 68 & B--60 & Si~III(1206) & 4637.42 & 2.8437 & 12.98$\pm$1.60 & 40 \cr
 69 & B--60 & Si~III(1206) & 4638.27 & 2.8444 & 13.20$\pm$0.76 & 38 \cr
 70 & B--60 & Si~III(1206) & 4639.23 & 2.8452 & 13.58$\pm$0.74 & 26 \cr
 71 & B--61 & \Lya & 4641.55 & 2.8181 & 14.36$\pm$0.38 & 45 \cr
\noalign{\vskip 6pt \hrule height .08em \vskip 6pt}
}
}$$
\vfill\eject
$$\vbox{
\halign{
\tabskip=2em
\hskip 2em \hfil # & # \hfil & # \hfil & \hfil#\hfil & \hfil#\hfil
& \hfil#\hfil &\hfil # \hfil \cr
\multispan{7}\hfil TABLE 5 --- {\it continued} \hfil \cr\cr
\noalign{ \hrule height .08em \vskip 2pt \hrule height .08em \vskip 6pt}
No. & \omit\hfil \rm Cross-ID \hfil  &  Line ID. & $\rm \lambda_{vac}^{a}$
& $\rm z_{abs}$ &  log $N$ (cm$^{-2}$) & $b$ (km~s$^{-1}$) \cr
\noalign{\vskip 6pt \hrule height .08em \vskip 6pt}
 72 & B--62 & \Lya & 4647.26 & 2.8228 & 13.63$\pm$0.59 & 33 \cr
 73 & B--62 & \Lya & 4648.48 & 2.8238 & 13.11$\pm$0.27 & 18 \cr
 74 &       & \Lya & 4650.79 & 2.8257 & 12.75$\pm$1.11 & 13 \cr
 75 & B--63 & \Lya & 4654.44 & 2.8287 & 13.88$\pm$0.56 & 26 \cr
 76 & B--63 & \Lya & 4654.92 & 2.8291 & 13.17$\pm$0.24 & 11 \cr
 77 &       & \Lya & 4655.89 & 2.8299 & 13.53$\pm$0.32 & 16 \cr
 78 &       & \Lya & 4657.84 & 2.8315 & 13.39$\pm$0.53 & 31 \cr
 79 & B--64 & \Lya & 4673.40 & 2.8443 & 20.23$\pm$0.41 &    \cr
 80 & B--65 & \Lya & 4692.37 & 2.8599 & 13.70$\pm$3.61 & 29 \cr
 81 & B--65 & \Lya & 4693.82 & 2.8611 & 13.03$\pm$4.34 & 18 \cr
 82 & B--66 & \Lya & 4697.11 & 2.8638 & 13.27$\pm$4.19 &  9 \cr
 83 & B--67 & \Lya & 4707.56 & 2.8724 & 13.48$\pm$0.74 & 49 \cr
 84 & B--68 & \Lya & 4714.49 & 2.8781 & 13.59$\pm$0.53 & 55 \cr
 85 & B--69 & \Lya & 4721.91 & 2.8842 & 14.29$\pm$0.34 & 80 \cr
 86 &       & \Lya & 4727.01 & 2.8884 & 13.37$\pm$0.46 & 35 \cr
 87 & B--70 & \Lya & 4731.14 & 2.8918 & 13.71$\pm$0.37 & 17 \cr
 88 & B--70 & \Lya & 4732.00 & 2.8925 & 14.10$\pm$0.35 & 30 \cr
 89 & B--70 & \Lya & 4733.09 & 2.8934 & 14.98$\pm$0.50 & 25 \cr
 90 & B--71 & \Lya & 4743.78 & 2.9022 & 13.56$\pm$0.41 & 56 \cr
 91 & B--72 & \Lya & 4749.62 & 2.9070 & 13.68$\pm$0.45 & 20 \cr
 92 & B--73 & \Lya & 4753.15 & 2.9099 & 13.92$\pm$0.34 & 27 \cr
 93 & B--74 & \Lya & 4758.74 & 2.9145 & 13.58$\pm$0.44 & 38 \cr
 94 & B--74 & \Lya & 4759.71 & 2.9153 & 13.66$\pm$0.48 & 23 \cr
 95 & B--74 & \Lya & 4761.29 & 2.9166 & 14.40$\pm$0.46 & 36 \cr
 96 & B--74 & \Lya & 4762.39 & 2.9175 & 14.16$\pm$0.50 & 10 \cr
 97 & B--74 & \Lya & 4762.75 & 2.9178 & 13.94$\pm$0.43 & 28 \cr
 98 & B--75 & \Lya & 4766.89 & 2.9212 & 13.67$\pm$0.38 & 43 \cr
 99 & B--76 & \Lya & 4774.42 & 2.9274 & 14.29$\pm$0.37 & 36 \cr
100 & B--78 & \Lya & 4793.75 & 2.9433 & 13.43$\pm$0.59 & 32 \cr
101 & B--78 & \Lya & 4794.60 & 2.9440 & 13.98$\pm$0.42 & 31 \cr
102 & B--82 & \Lya & 4824.63 & 2.9687 & 13.33$\pm$0.41 & 23 \cr
103 & B--83 & \Lya & 4831.32 & 2.9742 & 13.01$\pm$0.34 & 16 \cr
104 & B--84 & \Lya & 4836.06 & 2.9781 & 13.78$\pm$0.29 & 22 \cr
105 & B--85 & \Lya & 4842.74 & 2.9836 & 13.80$\pm$0.18 & 28 \cr
106 & B--85 & \Lya & 4843.96 & 2.9846 & 13.93$\pm$0.27 & 18 \cr
107 & B--85 & \Lya & 4844.93 & 2.9854 & 14.37$\pm$0.57 & 32 \cr
    &       & Si~II(1260)~ & 4844.94 & 2.8439 & 13.93$\pm$0.31 & 39 \cr
\noalign{\vskip 6pt \hrule height .08em \vskip 6pt}
}
}$$
\vfill\eject
$$\vbox{
\halign{
\tabskip=2em
\hskip 2em \hfil # & # \hfil & # \hfil & \hfil#\hfil & \hfil#\hfil
& \hfil#\hfil &\hfil # \hfil \cr
\multispan{7}\hfil TABLE 5 --- {\it continued} \hfil \cr\cr
\noalign{ \hrule height .08em \vskip 2pt \hrule height .08em \vskip 6pt}
No. & \omit\hfil \rm Cross-ID \hfil  &  Line ID. & $\rm \lambda_{vac}^{a}$
& $\rm z_{abs}$ &  log $N$ (cm$^{-2}$) & $b$ (km~s$^{-1}$) \cr
\noalign{\vskip 6pt \hrule height .08em \vskip 6pt}
108 & B--85 & \Lya & 4846.88 & 2.9870 & 13.75$\pm$0.56 & 38 \cr
    &       & Si~II(1260) & 4846.95 & 2.8455 & 13.37$\pm$0.45 & 39 \cr
109 & B--85 & \Lya & 4847.73 & 2.9877 & 12.62$\pm$3.81 & 17 \cr
110 & B--85 & \Lya & 4850.16 & 2.9897 & 13.40$\pm$0.32 & 29 \cr
111 & B--86 & \Lya & 4853.81 & 2.9927 & 14.09$\pm$0.24 & 38 \cr
112 & B--87 & \Lya & 4858.30 & 2.9964 & 13.53$\pm$0.26 & 30 \cr
113 &       & \Lya & 4865.35 & 3.0022 & 13.35$\pm$0.42 & 29 \cr
114 & B--89 & \Lya & 4871.55 & 3.0073 & 13.85$\pm$0.52 & 31 \cr
115 & B--89 & \Lya & 4873.62 & 3.0090 & 14.55$\pm$0.68 & 30 \cr
116 & B--90 & \Lya & 4880.55 & 3.0147 & 13.22$\pm$0.16 & 17 \cr
117 & B--90 & \Lya & 4881.94 & 3.0151 & 13.53$\pm$0.19 & 16 \cr
118 & B--90 & \Lya & 4882.01 & 3.0159 & 13.44$\pm$0.31 & 29 \cr
119 & B--91 & Si~III(1206) & 4885.36 & 3.0492 & 13.13$\pm$0.45 & 22 \cr
120 & B--91 & \Lya & 4886.26 & 3.0194 & 14.19$\pm$0.52 & 36 \cr
121 & B--91 & SI~III(1206) & 4886.81 & 3.0504 & 12.88$\pm$0.77 & 18 \cr
122 & B--92 & \Lya & 4889.42 & 3.0220 & 14.35$\pm$0.27 & 26 \cr
123 & B--92 & \Lya & 4890.40 & 3.0228 & 13.38$\pm$0.23 & 23 \cr
124 & B--93 & \Lya & 4895.38 & 3.0269 & 13.99$\pm$0.69 & 38 \cr
125 & B--93 & \Lya & 4896.72 & 3.0280 & 14.93$\pm$0.60 & 28 \cr
126 & B--93 & \Lya & 4898.91 & 3.0298 & 14.45$\pm$0.36 & 23 \cr
127 & B--95 & \Lya & 4909.48 & 3.0385 & 14.17$\pm$0.39 & 41 \cr
128 & B--96 & \Lya & 4917.63 & 3.0452 & 13.41$\pm$0.39 & 33 \cr
129 & B--96 & \Lya & 4919.21 & 3.0465 & 13.43$\pm$0.39 & 34 \cr
130 & B--97 & \Lya & 4921.76 & 3.0486 & 13.98$\pm$0.63 & 43 \cr
131 & B--97 & \Lya & 4922.61 & 3.0493 & 17.64$\pm$1.67 & 22 \cr
132 & B--97 & \Lya & 4924.07 & 3.0505 & 14.09$\pm$0.54 & 29 \cr
133 &       & \Lya & 4927.23 & 3.0531 & 13.12$\pm$0.27 & 24 \cr
134 & C--1  & O~I(1302) & 5005.93   & 2.8443 & 14.38$\pm$0.48 &  9 \cr
135 &       & Si~II(1304) & 5014.39 & 2.8443 & 13.10$\pm$0.57 &  9 \cr
136 & C--2  & N~V(1238) & 5016.85   & 3.0497 & 13.32$\pm$0.60 & 14 \cr
137 & C--2  & N~V(1238) & 5017.60   & 3.0503 & 12.90$\pm$1.21 & 12 \cr
138 & C--5  & Si~II(1260) & 5104.08 & 3.0495 & 12.58$\pm$0.61 & 14 \cr
139 & C--5  & Si~II(1260) & 5105.21 & 3.0504 & 12.57$\pm$0.59 & 11 \cr
140 & C--6  & C~II(1334)  & 5129.14 & 2.8434 & 13.34$\pm$0.43 &  7 \cr
141 & C--6  & C~II(1334)  & 5130.34 & 2.8443 & 13.96$\pm$0.47 & 11 \cr
142 &       & C~II$^*$(1335) & 5135.13 & 2.8445 & 13.22$\pm$0.68 & 11 \cr
\noalign{\vskip 6pt \hrule height .08em \vskip 6pt}
}
}$$
\vfill\eject
$$\vbox{
\halign{
\tabskip=2em
\hskip 2em \hfil # & # \hfil & # \hfil & \hfil#\hfil & \hfil#\hfil
& \hfil#\hfil &\hfil # \hfil \cr
\multispan{7}\hfil TABLE 5 --- {\it continued} \hfil \cr\cr
\noalign{ \hrule height .08em \vskip 2pt \hrule height .08em \vskip 6pt}
No. & \omit\hfil \rm Cross-ID \hfil  &  Line ID. & $\rm \lambda_{vac}^{a}$
& $\rm z_{abs}$ &  log $N$ (cm$^{-2}$) & $b$ (km~s$^{-1}$) \cr
\noalign{\vskip 6pt \hrule height .08em \vskip 6pt}
143 & C--8  & Si~IV(1402) & 5391.41 & 2.8434 & 13.39$\pm$0.47 & 14 \cr
144 & C--8  & Si~IV(1402) & 5392.11 & 2.8439 & 12.94$\pm$0.96 & 15 \cr
145 & C--8  & Si~IV(1402) & 5294.07 & 2.8453 & 13.03$\pm$0.73 & 14 \cr
146 & C--11 & Si~IV(1393) & 5642.34 & 3.0482 & 13.00$\pm$0.35 & 13 \cr
    & C--11 & C~IV(1548)  & 5642.27 & 2.6444 & 13.39$\pm$0.29 & 13 \cr
147 & C--11 & Si~IV(1393) & 5644.15 & 3.0496 & 13.97$\pm$0.28 & 38 \cr
148 & C--11 & Si~IV(1393) & 5645.56 & 3.0504 & 13.60$\pm$0.34 & 20 \cr
149 & C--11 & C~IV(1550)  & 5651.64 & 2.6444 & 13.49$\pm$0.34 & 13 \cr
150 & C--15 & Si~II(1526) & 5867.82 & 2.8434 & 12.40$\pm$0.78 & 6 \cr
151 & C--15 & Si~II(1526) & 5869.12 & 2.8443 & 12.76$\pm$0.67 & 10 \cr
152 & C--17 & Mg~II(2803)& 5940.82 & 1.1191 & 12.82$\pm$0.33 & 11\cr
153 & C--18 & C~IV(1548) & 5950.36 & 2.8434 & 13.73$\pm$0.01 & 17\cr
154 & C--18 & C~IV(1548) & 5950.98 & 2.8438 & 13.58$\pm$0.05 & 15 \cr
155 & C--18 & C~IV(1548) & 5953.15 & 2.8452 & 13.55$\pm$0.21 & 23 \cr
156 & C--19 & C~IV(1550) & 5960.24 & 2.8434 & 14.02$\pm$0.23 & 24 \cr
157 & C--19 & C~IV(1550) & 5960.86 & 2.8438 & 13.46$\pm$0.48 &  9 \cr
158 & C--19 & C~IV(1550) & 5963.04 & 2.8452 & 13.46$\pm$0.48 & 24 \cr
159 & C--20 & C~IV(1548) & 6267.59 & 3.0483 & 13.70$\pm$0.59 & 13 \cr
160 & C--20 & C~IV(1548) & 6269.29 & 3.0494 & 14.44$\pm$0.38 & 52 \cr
161 & C--20 & C~IV(1548) & 6270.84 & 3.0504 & 14.09$\pm$0.59 & 23 \cr
162 & C--21 & C~IV(1550) & 6278.00 & 3.0483 & 13.80$\pm$0.29 & 15 \cr
163 & C--21 & C~IV(1550) & 6279.86 & 3.0495 & 14.69$\pm$0.16 & 55 \cr
164 & C--21 & C~IV(1550) & 6281.26 & 3.0504 & 14.59$\pm$0.41 & 21 \cr
165 & D--5  & Fe~II(2344) & 6419.16 & 1.7383 & 13.99$\pm$0.34 & 16 \cr
166 &       & Al~II(1670) & 6421.50 & 2.8434 & 12.04$\pm$0.51 & 17 \cr
167 &       & Al~II(1670) & 6423.01 & 2.8443 & 12.22$\pm$0.32 & 10 \cr
168 &       & Fe~II(2374) & 6501.75 & 1.7382 & 14.21$\pm$0.41 & 16 \cr
169 & D--8  & Fe~II(2600) & 7119.79 & 1.7382 & 13.85$\pm$0.92 & 16 \cr
\noalign{\vskip 6pt \hrule height .08em \vskip 6pt}
}
}$$
\vfill\eject
\pageno=37
$$\vbox{
\halign{
\tabskip=2em
\hskip 2em \hfil#\hfil& \hfil #\hfil & #\hfil & \hfil#\hfil & \hfil # \hfil
& \hfil # \hfil \cr
\multispan{6}\hfil {TABLE 6}\hfil\cr
\multispan{6}\hfil LINES IN THE METAL SYSTEMS REDWARD OF \Lya EMISSION
\hfil \cr\cr
\noalign{ \hrule height .08em \vskip 2pt\hrule height .08em \vskip 6pt}
System &  $z_{abs}$ &  \omit\hfil Ion \hfil &  $\lambda_{\rm lab}$~(\AA)
& log $N$ cm$^{-2}$ & $b$ (km/s) \cr
\noalign{\vskip 6pt \hrule height .08em \vskip 6pt}
1  & 1.1190  & Mg~II & 2803.531 & 12.82$\pm$0.33 & 11 \cr
   &  \multispan{5}\hfil \hrulefill \hfil \cr
   &         & Mg~II & 2796.352 & \multispan{2} setup-C--\#16 \hfil \cr
   &         & Mg~II & 2803.531 & \multispan{2} setup-C--\#17 \hfil \cr
\multispan{6}\hfil \hrulefill \hfil \cr
2  & 1.7382  & Fe~II & 2344.214 & 13.99$\pm$0.34 & 16 \cr
   &         & Fe~II & 2374.461 & 14.21$\pm$0.41 & 16 \cr
   &         & Fe~II & 2600.173 & 13.85$\pm$0.92 & 16 \cr
   &  \multispan{5}\hfil \hrulefill \hfil \cr
   &         & Al~III& 1854.716 & \multispan{2} setup-C--\#4 \hfil \cr
   &         & Cr~II & 2056.254 & \multispan{2} setup-C--\#10 \hfil \cr
   &         & Cr~II & 2066.161 & \multispan{2} setup-C--\#13 \hfil \cr
   &         & Fe~II & 2344.214 & \multispan{2} setup-D--\#5 \hfil \cr
   &         & Fe~II & 2374.461 & \multispan{2} setup-D--\#6 \hfil \cr
   &         & Fe~II & 2382.756 & \multispan{2} setup-D--\#7 \hfil \cr
   &         & Fe~II & 2586.650 & \multispan{2} setup-D--\#8 \hfil \cr
   &         & Fe~II & 2600.173 & \multispan{2} setup-D--\#9 \hfil \cr
   &         & Mg~II & 2796.352 & \multispan{2} setup-D--\#13 \hfil \cr
   &         & Mg~II & 2803.531 & \multispan{2} setup-D--\#14 \hfil \cr
\multispan{6}\hfil \hrulefill \hfil \cr
3  & 2.6444  & H~I   & 1215.670 & 17.37$\pm$0.56 &  7 \cr
   &         & C~IV  & 1548.202 & 13.49$\pm$0.29 & 13 \cr
   &         & C~IV  & 1550.774 & 13.49$\pm$0.34 & 13 \cr
   &  \multispan{5}\hfil \hrulefill \hfil \cr
   &         & C~IV  & 1550.770 & \multispan{2} setup-C--\#12 \hfil \cr
\noalign{\vskip 6pt \hrule height .08em \vskip 6pt}
4  & 2.844   & \multispan{4} \hfil 4-component absorption complex \hfil \cr
\noalign{\vskip 6pt \hrule height .08em \vskip 6pt}
   &         & O~I   & 1302.169 & \multispan{2} setup-C--\#1 \hfil \cr
   &         & C~II  & 1334.532 & \multispan{2} setup-C--\#6 \hfil \cr
   &         & Si~IV & 1393.755 & \multispan{2} setup-C--\#7 \hfil \cr
   &         & Si~IV & 1402.770 & \multispan{2} setup-C--\#8 \hfil \cr
   &         & Si~II & 1526.707 & \multispan{2} setup-C--\#15 \hfil \cr
   &         & C~IV  & 1548.195 & \multispan{2} setup-C--\#18 \hfil \cr
   &         & C~IV  & 1550.770 & \multispan{2} setup-C--\#19 \hfil \cr
\noalign{\vskip 6pt \hrule height .08em \vskip 6pt}
}
}$$
\vfill\eject
$$\vbox{
\halign{
\tabskip=2em
\hskip 2em \hfil#\hfil& \hfil #\hfil & #\hfil & \hfil#\hfil & \hfil # \hfil
& \hfil # \hfil \cr
\multispan{6}\hfil {TABLE 6 {\it continued}}\hfil\cr\cr
\noalign{ \hrule height .08em \vskip 2pt\hrule height .08em \vskip 6pt}
System &  $z_{abs}$ &  \omit\hfil Ion \hfil &  $\lambda_{\rm lab}$~(\AA)
& log $N$ cm$^{-2}$ & $b$ (km/s) \cr
\noalign{\vskip 6pt \hrule height .08em \vskip 6pt}
4a & 2.8434  & C~II  & 1334.532 & 13.34$\pm$0.43 &  7 \cr
   &         & Si~IV & 1402.770 & 13.39$\pm$0.47 & 14 \cr
   &         & Si~II & 1526.707 & 12.40$\pm$0.78 &  6 \cr
   &         & C~IV  & 1548.202 & 13.73$\pm$0.01 & 17 \cr
   &         & C~IV  & 1550.774 & 14.02$\pm$0.23 & 24 \cr
   &         & Al~II & 1670.787 & 12.04$\pm$0.51 & 17 \cr
4b & 2.8438  & Si~IV & 1402.770 & 12.39$\pm$0.96 & 15 \cr
   &         & C~IV  & 1548.202 & 13.58$\pm$0.05 & 15 \cr
   &         & C~IV  & 1550.774 & 13.46$\pm$0.48 &  9 \cr
4c & 2.8443  & H~I   & 1215.670 & 20.23$\pm$0.41 &    \cr
   &         & O~I   & 1302.169 & 14.38$\pm$0.48 &  9 \cr
   &         & Si~II & 1304.370 & 13.10$\pm$0.57 &  9 \cr
   &         & C~II  & 1334.532 & 13.96$\pm$0.47 & 11 \cr
   &      & C~II$^*$ & 1335.708 & 13.22$\pm$0.68 & 11 \cr
   &         & Si~II & 1526.707 & 12.76$\pm$0.67 & 10 \cr
   &         & Al~II & 1670.787 & 12.22$\pm$0.32 & 10 \cr
4d & 2.8452  & Si~IV & 1402.770 & 13.03$\pm$0.73 & 14 \cr
   &         & C~IV  & 1548.202 & 13.55$\pm$0.21 & 23 \cr
   &         & C~IV  & 1550.774 & 13.46$\pm$0.48 & 24 \cr
\noalign{\vskip 6pt \hrule height .08em \vskip 6pt}
5  & 2.8927  & H~I   & 1215.670 & 14.10$\pm$0.35 & 30 \cr
   &  \multispan{5}\hfil \hrulefill \hfil \cr
   &         & C~IV  & 1548.195 & \multispan{2} setup-C--\#20 \hfil \cr
   &         & C~IV  & 1550.770 & \multispan{2} setup-C--\#21 \hfil \cr
\noalign{\vskip 6pt \hrule height .08em \vskip 6pt}
6  & 3.050   &\multispan{4} \hfil 3-component absorption complex \hfil \cr
\noalign{\vskip 6pt \hrule height .08em \vskip 6pt}
   &         & N~V   & 1238.821 & \multispan{2} setup-C--\#2 \hfil \cr
   &         & N~V   & 1242.804 & \multispan{2} setup-C--\#3 \hfil \cr
   &         & Si~II & 1260.422 & \multispan{2} setup-C--\#5 \hfil \cr
   &         & C~II  & 1334.532 & \multispan{2} setup-C--\#9 \hfil \cr
   &         & Si~IV & 1393.755 & \multispan{2} setup-C--\#11 \hfil \cr
   &         & Si~IV & 1402.770 & \multispan{2} setup-C--\#14 \hfil \cr
   &         & C~IV  & 1548.195 & \multispan{2} setup-D--\#3 \hfil \cr
   &         & C~IV  & 1550.770 & \multispan{2} setup-D--\#4 \hfil \cr
   &  \multispan{5}\hfil \hrulefill \hfil \cr
6a & 3.0483  & H~I   & 1215.670 & 13.98$\pm$0.63 & 43 \cr
   &         & Si~IV & 1393.755 & 13.00$\pm$0.35 & 13 \cr
   &         & C~IV  & 1548.202 & 13.70$\pm$0.59 & 13 \cr
   &         & C~IV  & 1550.774 & 13.80$\pm$0.29 & 15 \cr
\noalign{\vskip 6pt \hrule height .08em \vskip 6pt}
}
}$$
\vfill\eject
$$\vbox{
\halign{
\tabskip=2em
\hskip 2em \hfil#\hfil& \hfil #\hfil & #\hfil & \hfil#\hfil & \hfil # \hfil
& \hfil # \hfil \cr
\multispan{6}\hfil {TABLE 6 {\it continued}}\hfil\cr\cr
\noalign{ \hrule height .08em \vskip 2pt\hrule height .08em \vskip 6pt}
System &  $z_{abs}$ &  \omit\hfil Ion \hfil &  $\lambda_{\rm lab}$~(\AA)
& log $N$ cm$^{-2}$ & $b$ (km/s) \cr
\noalign{\vskip 6pt \hrule height .08em \vskip 6pt}
6b & 3.0495  & H~I   & 1215.670 & 17.64$\pm$1.67 & 22 \cr
   &         & N~V   & 1238.821 & 13.32$\pm$0.60 & 14 \cr
   &         & Si~II & 1260.422 & 12.58$\pm$0.61 & 14 \cr
   &         & Si~IV & 1393.755 & 13.97$\pm$0.28 & 38 \cr
   &         & C~IV  & 1548.202 & 14.44$\pm$0.38 & 52 \cr
   &         & C~IV  & 1550.774 & 14.69$\pm$0.16 & 55 \cr
6c & 3.0504  & H~I   & 1215.670 & 14.09$\pm$0.54 & 29 \cr
   &         & N~V   & 1238.821 & 12.90$\pm$1.21 & 12 \cr
   &         & Si~II & 1260.422 & 12.57$\pm$0.59 & 11 \cr
   &         & Si~IV & 1393.755 & 13.60$\pm$0.34 & 20 \cr
   &         & C~IV  & 1548.202 & 14.09$\pm$0.59 & 23 \cr
   &         & C~IV  & 1550.774 & 14.59$\pm$0.41 & 21 \cr
\noalign{\vskip 6pt \hrule height .08em \vskip 6pt}
}
}$$
\vfill\eject
\pageno=40
$$\vbox{
\halign{
\tabskip=2em
\hskip 2em #\hfil & \hfil#
                  & \hfil# \hfil
                  & \hfil# \hfil
                  & \hfil#
                  & \hfil# \hfil \cr
\multispan{6}\hfil {TABLE 7}\hfil\cr
\multispan{6}\hfil COLUMN DENSITIES MEASURED FOR \hfil \cr
\multispan{6}\hfil THE ABSORPTION SYSTEM AT $z_{abs}$=2.844 \hfil\cr\cr
\noalign{ \hrule height .08em \vskip 2pt\hrule height .08em \vskip 6pt}
\omit\hfil Ion \hfil & \omit\hfil $\lambda_{\rm lab}$ \hfil
& log $N$ (cm$^{-2}$) & $b$ (km s$^{-1}$) &\omit\hfil $\tau_0$\hfil
& $z_{abs}$ \cr
\noalign{\vskip 6pt \hrule height .08em \vskip 6pt}
\multispan{6}\hfil Voigt Profile Fit to Only the 2.8443 Component \hfil \cr
\noalign{\vskip 6pt \hrule height .08em \vskip 6pt}
H~I~(Damped \Lya)& 1215.670 & 20.23$\pm$0.41 &   & $\sim > 10^3$ & 2.8443 \cr
O~I         & 1302.169 & 14.38$\pm$0.48 &  9  &   2.54  & 2.8443 \cr
C~II        & 1334.532 & 13.96$\pm$0.47 & 11  &   2.12  & 2.8443 \cr
C~II$^*$    & 1335.708 & 13.22$\pm$0.68 & 11  &   0.35  & 2.8445 \cr
C~IV~$^a$   &          & 13.02          & 10  &   0.42  & 2.8443 \cr
Al~II       & 1670.787 & 12.22$\pm$0.32 & 10  &   0.76  & 2.8443 \cr
Si~II       & 1190.416 & 13.60$\pm$0.60 & 10  &   1.78  & 2.8443 \cr
Si~II       & 1193.290 & 13.28$\pm$0.45 & 12  &   1.42  & 2.8441 \cr
Si~II       & 1304.416 & 13.10$\pm$0.57 &  9  &   0.40  & 2.8443 \cr
Si~II       & 1526.707 & 12.76$\pm$0.67 & 10  &   0.30  & 2.8443 \cr
Si~III~$^b$ & 1206.500 & 13.20$\pm$0.76 & 38  &   0.78  & 2.8444 \cr
Si~IV~$^a$  & 1402.770 & 12.58          & 10        &   0.19  & 2.8443 \cr
\noalign{\vskip 6pt \hrule height .08em \vskip 6pt}
\multispan{6}\hfil Curve Of Growth Fit to Whole $z_{abs}$=2.844 Complex
\hfil\cr
\noalign{\vskip 6pt \hrule height .08em \vskip 6pt}
\multispan{2}H~I Lyman series lines \hfil & 19.8$\pm$0.4 & 35$\pm$5 &&
2.8443\cr
O~I         & 1302.169 & 14.1$\pm$0.1 & 41$\pm$5~$^c$ && 2.8447 \cr
C~II        & 1334.532 & 14.0$\pm$0.1 & 41$\pm$5~$^c$ && 2.8455 \cr
C~III       &  977.020 & 16.2$\pm$0.7 & 41$\pm$5~$^{c}$ && 2.8442 \cr
C~IV        & 1548.202 & 14.3$\pm$0.1 & 41$\pm$5      && 2.8440 \cr
            & 1550.770 &              &               && 2.8439 \cr
Si~II       & 1526.707 & 13.3$\pm$0.1 & 41$\pm$5~$^c$ && 2.8446 \cr
Si~IV       & 1393.755 & 13.9$\pm$0.1 & 41$\pm$5      && 2.8442 \cr
            & 1402.770 &              &               && 2.8442 \cr
\noalign{\vskip 6pt \hrule height .08em \vskip 6pt}
\noalign{\hbox{$^a$ From 5 pixels equivalent width on the linear part of
curve of growth.} \vskip 6pt}
\noalign{\hbox{$^b$ Blended with other absorption lines.}\vskip 6pt}
\noalign{\hbox{$^c$ Mean $b$ used to get column density.}}
}
}$$
\vfill\eject
\pageno=41
$$\vbox{
\halign{
\tabskip=2em
\hskip 2em  # \hfil & \hfil#\hfil & \hfil # \hfil & \hfil # \hfil \cr
\multispan{4}\hfil {TABLE 8}\hfil\cr
\multispan{4}\hfil COLUMN DENSITY RATIOS IN THE $z_{abs}=2.8443$ DLA SYSTEM
\hfil \cr\cr
\noalign{ \hrule height .08em \vskip 2pt\hrule height .08em \vskip 6pt}
     & Observed & $\Gamma~^{a}$  & Ratio At  \cr
     &   Ratio  & From Ratio     & $\Gamma=-2.75~^b$ \cr
\noalign{\vskip 6pt \hrule height .08em \vskip 6pt}
log $N$(C~II)/$N$(C~IV)   & 0.92 & --2.68 & 1.07 \cr
log $N$(C~II)/$N$(Si~IV)  & 1.38 & --2.57 & 1.65 \cr
log $N$(C~IV)/$N$(Si~IV)  & 0.44 & --3.09 & 0.58 \cr
log $N$(C~IV)/$N$(Al~II)  & 0.80 & --2.77 & 0.76 \cr
log $N$(O~I)/$N$(C~IV)    & 1.36 & --2.79 & 1.29 \cr
log $N$(O~I)/$N$(Si~IV)   & 1.80 & --2.71 & 1.87 \cr
log $N$(Si~II)/$N$(C~IV)  & 0.18 & --2.80 & 0.08 \cr
log $N$(Si~II)/$N$(Si~IV) & 0.62 & --2.72 & 0.66 \cr
log $N$(Si~IV)/$N$(Al~II) & 0.36 & --2.66 & 0.18 \cr
&      & $\langle \Gamma \rangle$=--2.75  &      \cr
\noalign{\vskip 6pt \hrule height .08em \vskip 6pt}
\noalign{\hbox{$^a$ $\Gamma$ value where the calculated column density ratio
matches the observed value.} \vskip 6pt}
\noalign{\hbox{$^b$ Column density ratio calculated from CLOUDY code with
$\Gamma$=--2.75,} \vskip 6pt}
\noalign{\hbox{log $N$(H~I)=20.2 and [M/H]=--2.8.} \vskip 6pt}
}
}$$
\bigskip\bigskip
$$\vbox{
\halign{
\tabskip=2em
\hskip 2em \hfil# \hfil & \hfil#\hfil & \hfil#\hfil  & \hfil#\hfil
& \hfil#\hfil\cr
\multispan{5}\hfil TABLE 9  \hfil\cr
\multispan{5}\hfil ABUNDANCE RATIOS IN THE $z_{abs}=2.8443$ DLA SYSTEM
\hfil \cr\cr
\noalign{ \hrule height .08em \vskip 2pt \hrule height .08em \vskip 6pt}
Ion & \multispan{4} \hfil ABUNDANCE RELATIVE TO COSMIC \hfil \cr
    & \multispan{4} \hrulefill \cr
 X   & [X/C]  & [X/O] & [X/Al] & [X/Si] \cr
\noalign{\vskip 6pt \hrule height .08em \vskip 6pt}
     & \multispan{4}\hfil  DLA SYSTEM AT $z_{abs}=2.8443$ \hfil \cr
C   &    ...        &0.06(--0.2)~$^a$ & --0.23(--0.1) & --0.23(--0.2) \cr
O   & --0.06(+0.2)  &   ...           & --0.29(+0.1)  & --0.29( 0.0)  \cr
Al  &   0.23(+0.1)  &   0.29(--0.1)   &   ...         &   0.00(-0.1)  \cr
Si  &   0.23(--0.2) &   0.29(  0.0)   &   0.00(+0.1)  &   ...         \cr
\noalign{\vskip 6pt \hrule height .08em \vskip 6pt}
     & \multispan{4}\hfil  DEPLETED ISM ($\alpha$ Sco A) GAS$^b$ \hfil \cr
C   &  ...  &   0.02 &   1.11 &   0.12 \cr
O   &--0.02 &   ...  &   1.09 &   0.10 \cr
Al  &--1.11 & --1.09 &   ...  & --0.99 \cr
Si  &--0.12 & --0.10 &   0.99 &   ...  \cr
\noalign{\vskip 6pt \hrule height .08em \vskip 6pt}
\noalign{\hbox{$^a$ The leading values are for observed ions only
whereas those in parenthesis are}\vskip 6pt}
\noalign{\hbox{corredted for ionization} \vskip 6pt}
\noalign{\hbox{$^b$ From Jenkins (1987).}\vskip 6pt}
}
}$$
\vfill\eject
\pageno=42
\halign{
\tabskip=1em
\hskip 1em #\hfil & \hfil#\hfil & \hfil#\hfil & \hfil#\hfil & \hfil#\hfil
    & \hfil#\hfil
    & \hfil#\hfil & \hfil#\hfil & \hfil#\hfil \cr
\multispan{9}\hfil TABLE 10  \hfil\cr
\multispan{9}\hfil STATISTICS OF \Lya ABSORPTION LINES \hfil \cr\cr
\noalign{ \hrule height .08em \vskip 2pt \hrule height .08em \vskip 6pt}
\omit\hfil QSO \hfil & FWHM &  N & $\langle z_{abs} \rangle$
                         & $\langle b \rangle$
                         & $\langle {\rm log} N({\rm HI}) \rangle$
                         & \multispan{2}\hfil Fraction with $b \leq 20$\kms
\hfil
                         & Ref.~$^a$ \cr
 & (\kms) & & & & & log $N$(H~I)$\le$13.5
                         & log $N$(H~I)$>$13.5  & \cr
\noalign{\vskip 6pt \hrule height .08em \vskip 6pt}
\multispan{9}\hfil All \Lya lines without metals \hfil \cr
0014+8118    & 20    & 294 & 3.0248 & 32 & 13.63 &40/140~$^b$=28.6\% &
10/154=6.5\% & 1 \cr
1100$-$264   &8.5--9 &  71 & 1.9895 & 34 & 13.48 &  9/40=22.5\% &  3/31= 9.7\%
& 2 \cr
1946+7658    & 10  & 113 & 2.7926 & 27 & 13.85 & 12/28=42.8\% & 28/85=32.9\% &
3 \cr
1946+7658~$^c$& 10  &  43 & 2.9674 & 31 & 13.78 &  3/14=21.4\% &  4/29=13.8\% &
3 \cr
2126$-$158   & 14    & 130 & 3.0782 & 27 & 13.62 & 35/65=53.8\% & 13/65=20\%
& 4 \cr
2206$-$199N  &  6    & 101 & 2.3501 & 27 & 13.54 & 30/60=50.0\% &  2/41= 4.9\%
& 6 \cr
\noalign{\vskip 6pt \hrule height .08em \vskip 6pt}
\multispan{9}\hfil Unsaturated \Lya lines without metals \hfil \cr
1946+7658    & 10  &  96 & 2.7976 & 28 & 13.64 & 12/28=42.8\% & 19/68=27.9\% &
3 \cr
1946+7658~$^c$& 10  &  36 & 2.9661 & 32 & 13.63 &  3/14=21.4\% &  4/22=18.2\% &
3 \cr
2126$-$158   & 14    &  93 & 3.0779 & 27 & 13.43 & 34/56=60.1\% &  4/37=10.8\%
& 4 \cr
2206$-$199N  &  6    &  41 & 2.3721 & 17 & 13.34 & 24/27=88.9\% &  4/14=28.6\%
& 5 \cr
2206$-$199N  &  6    &  39 & 2.3733 & 22 & 13.37 & 17/26=65.4\% &  1/13=7.8\%
& 6 \cr
\noalign{\vskip 6pt \hrule height .08em \vskip 6pt}
\noalign{\hbox{$^a$ Ref. (1) Rauch \etal (1992); (2) Carswell \etal (1991); (3)
This paper;
(4) Giallongo \etal (1993); } \vskip 6pt}
\noalign{\hbox{(5) Pettini \etal (1990a); (6) Rauch \etal (1993).} \vskip 6pt}
\noalign{\hbox{$^b$ 40 out of 140 lines with log $N$(H~I) $\leq 13.5$ have
$b\leq 20$.} \vskip 6pt}
\noalign{\hbox{$^c$ Ly$\alpha$ lines in spectra with 8 $<$ SNR $<$ 16,
which are all at $\lambda \geq 4700$~\AA.}\vskip 6pt}
}
%
\vfill\eject
\pageno=43
\baselineskip=0.6cm
\main
%
%
\centerline{REFERENCES}
\medskip
\hangpara Atwood, B., Baldwin, J. A., \& Carswell, R. F. 1985, ApJ, 292, 58
\hangpara Babul, A. 1991, MNRAS, 248, 177
\hangpara Bajtlik, S., Duncan, R. C., \& Ostriker, J. P. 1988, ApJ, 327, 570
\hangpara Bahcall, J. N., \& Wolf, R. A. 1968, ApJ, 152, 701
\hangpara Bechtold, J., Weymann, R. J., Zuo, L., \& Malkan, M. A. 1987,
          ApJ, 281, 76
\hangpara Bechtold, J. 1993, preprint.
\hangpara Carswell, R. F., \& Rees, M. J. 1987, MNRAS, 224, 13
\hangpara Carswell, R. F., Webb, J. K., Baldwin, J. A., \& Atwood, B. 1987,
          ApJ, 319, 709
\hangpara Carswell, R. F., Lanzetta, K. M., Parnell, H. C., \& Webb, J. K.
          1991, ApJ, 371, 36
\hangpara Fan, F. X., \& Tytler, D. 1989, BAAS, 21, 797
\hangpara Fall, S. M., \& Pei, Y. C. 1989, ApJ, 337, 7
\hangpara Ferland, G. J. 1991, Ohio State University (OSU)
          Astronomy Department Internal Report 91--01.
\hangpara Giallongo, E., Cristiani, S., Fontana, A., \& Tr\`evese, D. 1993,
          ApJ, 416, 137
\hangpara Hagen, H.-J., Cordis, L., Engels, D., Groote, D., Haug, U.,
          Heber, U., K\"ohler, Th., Wisotzki, L., \& Reimers, D. 1992,
          A\&A, 253, L5
\hangpara Horne, K. 1986, PASP, 98, 609
\hangpara Jenkins, E. B. 1986, ApJ, 304, 739
\hangpara Jenkins, E. B. 1987, in {\sl  Interstellar Processes}, p533,
          eds. D. J. Hollenbach \& H. A. Thronson, Jr.
\hangpara James, F., \& Roos, M. 1975, Comput. Phys. Commun. 10, 343
\hangpara Lanzetta, K. M., \& Bowen, D. V. 1992, ApJ, 391, 48
\hangpara Lanzetta, K. M., Bowen, D. V., Tytler, D., \& Webb, J. K. 1994,
          ApJ, submitted.
\hangpara Lu, L., Wolfe, A. M., \& Turnshek, D. A. 1991, ApJ, 367, 19
\hangpara Lu, L., Wolfe, A. M., Turnshek, D. A., \& Lanzetta, K. M.
          1993, ApJS, 84, 1
\hangpara Miller, J. S., \& Stone, R. P. S. 1987, Lick Observatory Technical
          Report 48, {\sl  The CCD Cassegrain Spectrograph at the Shane
          Reflector}
\hangpara Misch, A. 1991, Lick Observatory Technical Report 58,
          {\sl  The Hamilton Spectrograph User's Manual}.
\hangpara Morris, S. L., Weymann, R. J., Foltz, C. B., Turnshek, D. A.,
          Shectman, S., Price, S., \& Boroson, T. A., 1986, ApJ, 310, 40
\hangpara Murdoch, H. S., Hunstead, R. W., Peterson, B. A., Blades, J. C.,
          Jauncey, D. L., Wright, A. E., Pettini, M., \& Savage, A.
          1986, ApJ, 305, 496
\hangpara Ostriker, J. P., Bajtlik, S. \& Duncan, R. C. 1988, ApJL, 327, L35
\hangpara Osterbrock, D. E., \& Martel, Andr\'e 1992, PASP, 104, 76
\hangpara Pettini, M., Hunstead, R. W., Smith, L. J., \& Mar, D. P. 1990a,
          MNRAS, 246, 545 (PHSM)
\hangpara Pettini, M., Boksenberg, A., \& Hunstead, R. W. 1990b,
          ApJ, 348, 48
\hangpara Pettini, M., Smith, L. J., Hunstead, R. W., \& King, D. L. 1994,
          ApJ, submitted.
\hangpara Rauch, M., Carswell, R. F., Robertson, J. G., Shaver, P. A.,
          \& Webb, J. K. 1990, MNRAS, 242, 698
\hangpara Rauch, M., Carswell, R. F., Chaffee, F. H, Foltz, C. B.,
          Webb, J. K., Weymann, R. J., Bechtold, J., \& Green, R. F.
          1992, ApJ, 390, 387
\hangpara Rauch, M., Carswell, R. F., Webb, J. K., \& Weymann, R. J.
          1993, MNRAS, 260, 589
\hangpara Reimers, D., Vogel, S., Hagen, H.-J., Engels, D., Groote, D.,
          Wamsteker, W., Clavel, J. \& Rosa, M. R. 1992, Nature, 360, 561
\hangpara Sanz, J. L., Clavel, J., Naylor, T., \& Wamsteker, W. 1993,
          MNRAS, 260, 468
\hangpara Sargent, W. L. W., Young, P., Boksenberg, A., \& Tytler, D.
          1980, ApJS, 42, 41
\hangpara Sargent, W. L. W., Steidel, C. C., \& Boksenberg, A. 1990,
          ApJ, 351, 364
\hangpara Spitzer, Jr., L. 1978, {\sl  Physical processes in the interstellar
          medium}
\hangpara Steidel, C. C. 1990, ApJS, 74, 37
\hangpara Turnshek, D. A., Wolfe, A. M., Lanzetta, K. M., Briggs, F. H.,
          Cohen, R. D., Foltz, C. B., Smith, H. E. and Wilkes, B. J.
          1989, ApJ, 344, 567
\hangpara Tytler, D. 1982, Ph.D. Thesis, University of College London.
\hangpara Tytler, D. 1987a, ApJ, 321, 69
\hangpara Tytler, D., \& Fan, X.-M. 1992, ApJS, 79, 1
\hangpara Tytler, D., Fan, X.-M., Junkkarinen, V. T., \& Cohen, R. S.
          AJ, 1993, 106, 426
\hangpara Vogt, S. S. 1987, PASP, 99, 1214
\hangpara Vogel, S., \& Reimers, D. 1993 A\&A, 274, L5
\hangpara Webb, J. K. \& Barcons, X. 1991, MNRAS, 250, 270
\hangpara Wolfe, A. M., Turnshek, D. A., Lanzetta, K. M., \& Lu, L.
          1993, ApJ, 404, 480
\hangpara Young, P. J., Sargent, W. L. W, Boksenberg, A., Carswell, R. F.,
          \& Whelan, J. A. 1979, ApJ, 229, 891

\vfill\eject
%
%
\centerline{FIGURE CAPTIONS}
\medskip
\item{Fig. 1} Intermediate resolution spectra of HS~1946+7658,
along with the 1~$\sigma$ noise array: (a) setup-A; (b) setup-B;
(c) setup-C; (d) setup-D and (e) setup-E. The flux is $f_{\nu}$, in
units of micro-Jansky. We did not smooth the data, and we show
individual pixels. Arrows show the emission line peak positions which
we list in Table 3.

\medskip
\item{Fig. 2} Echelle spectrum of HS~1946+7658, normalized to unit
continuum ({\it histogram}), and fitted Voigt profiles
({\it smoothed solid line}). We did not smooth the data, and we show
each $2 \times 2$ binned pixel ($\sim$ 0.09~\AA).

\medskip
\item{Fig. 3} Metal absorption lines on a velocity scale relative to
the low ionization component at $z_{abs}=$2.8443.  The vertical
{\it dash} lines show the 4 absorption components, from left to
right, $z_{abs}$=2.8434, 2.8438, 2.8443, 2.8452. The echelle
data ({\it histogram}) have been fitted with Voigt profiles
({\it smooth solid line}).
Each line has been normalized to unit continuum intensity level.
We fitted each line independently, so they need not have identical
velocities. Low ionization lines are concentrated in the
$z_{abs}$=2.8443 component, while the high ionization lines are in
other components.

\medskip
\item{Fig. 4} A Voigt profile fit to H Lyman series lines of the DLA system
at $z_{abs}=2.8443$ with log $N$(H~I)=19.8 and $b$=35 \kms.
The {\it histogram} shows the observed data and the {\it dotted line}
shows the fit. The resulting simulated spectrum was convolved with the
instrumental profile for a direct comparison with the observation.
The LLS at \L $\sim$ 3520~\AA\ is caused
by the DLA system. Note that the neutral and singly ionized ions have
$b$=10 \kms, so some H~I is extended in velocity space.

\medskip
\item{Fig. 5} Metal absorption lines in the $z_{abs}$=3.050 complex shown on
a velocity scale relative to $z_{abs}$=3.050.
The complex has three velocity components, indicated by the $dash$
vertical lines, which correspond to (left to right)
$z_{abs}$=3.0482, 3.0495 and  3.0504.
The echelle spectra ({\it histogram}) have been fitted with Voigt
profiles ({\it smooth solid line}). Each segment has been normalized to unit
continuum.

\medskip
\item{Fig. 6} Curve of growth for the $z_{abs}$=2.844 system.
The {\it solid line} is a $b$=41 \kms best fit to metal lines
redward of \Lya emission line. The {\it dashed} line has
$b_{\rm H}$=35 \kms and is a better fit to the
H Lyman series lines which are represented by stars, from right
to left \Lya, \Lyb, \Lyg, etc. Damping wings are important at
$log(Nf\lambda ) \geq 12$.
In the upper right corner we show curves for damped \Lya
(upper) and damped \Lyb (lower) only. In each case the {\it
dot-dash} line is for $b=41$, while the {\it dotted} line is
for $b=35$. \Lyg and \Lyd are blends.

\medskip
\item{Fig. 7}  The C~II\L1334 and  C~II$^*$\L1335 lines
on a velocity scale with respect to $z_{abs}$=2.8443.
The {\it Histogram} and the {\it smooth solid line} represent the data and
fitted profile respectively.  There is a 15 \kms velocity
shift from C~II to C~II$^*$. On the bottom panel
we show the column density ratio $R(v)$ (see \S4.2.3).

\medskip
\item{Fig. 8}
Photoionization model results for a cloud with column density
log~$N$(H~I)=20.2, and metal abundance [M/H]=$-$2.8. The plot
shows the expected relative column
densities for ions of carbon, silicon, oxygen and aluminum, as a function of
ionization parameter $\Gamma \equiv n_{\gamma}/n_{\rm H}$.
The names of the ions are written above their lines, and
we show the measurements of the column densities of
O~I ({\it cross}),
C~II ({\it open triangle}),
Si~II ({\it open diamond}),
C~IV ({\it filled circle}),
Si~IV ({\it filled square})
and Al~II ({\it open square}).
We adopt log~$\Gamma=-2.75\pm0.30$.

\medskip
\item{Fig. 9}  Distribution of \Lya lines with W$_{r} \geq 0.36$~\AA\ from the
spectra of 16 QSOs all with DLA systems.
The co-evolving coordinate is $\chi_{\gamma}$, with $\gamma=2.7$,
where $\chi_{\gamma}=[(1+z)^{\gamma}-1]/(1+\gamma)$.
Each {\it vertical line} indicates a \Lya line, and the DLA
line regions are indicated by the {\it horizontal lines}.
The mean line density does appear to very a lot on scales of
$\chi \simeq 5$, but HS~1946+769 is typical,
and there is no general change in the
\Lya line density near the DLA systems.

\medskip
\item{Fig. 10}  Distribution of Doppler parameter $b$ versus log $N$(H~I)
for the \Lya forest absorption lines: {\it solid dots} and {\it crosses}
represent those unsaturated lines with SNR$<$8 (at \L $<$ 4500~\AA)
and SNR$>8$ (at \L $>$ 4700~\AA) respectively; {\it open circles}
represent saturated lines and {\it pluses} indicate the saturated lines with
SNR$>$8 . The theoretical curves show $b$ against
log $N$(H~I) for  rest equivalent widths
W$_r$=0.40~\AA\ ({\it solid line}),
W$_r$=0.32~\AA\ ({\it dashed line}),
W$_r$=0.20~\AA\ ({\it dot-dash-dot line}),
W$_r$=0.10~\AA\ ({\it dotted line}),
W$_r$=0.05~\AA\ ({\it dash-dot-dot-dot line}).

\bye